\documentclass[
aps,
prd,
10pt,
twocolumn,
nofootinbib,
preprintnumbers,
superscriptaddress,
showpacs]{revtex4-2}

\usepackage{comment}
\usepackage{booktabs,makecell}
\usepackage{graphicx}
\usepackage{amsmath,amssymb}
\usepackage{amsfonts}
\usepackage{xspace} 
\usepackage[usenames]{color}
\usepackage{dcolumn}
\usepackage{bm}
\usepackage{mathrsfs}
\usepackage[colorlinks=true,citecolor=blue]{hyperref}
\usepackage[all]{hypcap} 
\usepackage[utf8]{inputenc} 
\usepackage{slashed}
\usepackage{multirow}
\usepackage{rotating}
\definecolor{orange}{rgb}{1,0.5,0}
\usepackage{tabularx}
\usepackage{siunitx}
\usepackage[normalem]{ulem}
\usepackage[dvipsnames]{xcolor}

\def\dd{{\rm d}}

\def\rrBM{\ifmmode r^{\text{BM}}_{\text{ratio}} \else {$r^{\text{BM}}_{\text{ratio}}$}\fi}
\def\rrDM{\ifmmode r^{\text{DM}}_{\text{ratio}} \else {$r^{\text{DM}}_{\text{ratio}}$}\fi}
\def\X{{\text{X}}}
\def\K{{\text{K}}}
\def\BM{{\text{BM}}}
\def\DM{{\text{DM}}}
\def\Msun{{\text{\,M}_\odot}}

\def\tot{{\text{tot}}}
\def\max{{\text{max}}}
\def\fm{{f^M_\DM}}
\def\fj{{f^J_\DM}}

\DeclareSIUnit{\solarmass}{\text{M}_\odot}
\DeclareSIUnit{\clight}{\text{c}}
\DeclareSIUnit{\Junit}{\text{G}\,\text{M}_\odot^2\,\text{c}^{-1}}
\DeclareSIUnit{\dens}{\gram\per\centi\meter\cubed}

\begin{document}

\title{Differentially rotating neutron stars with dark matter cores}

\author{Lorenzo Cipriani}\email{lorenzo.cipriani@graduate.univaq.it}
\affiliation{Dipartimento di Scienze Fisiche e Chimiche, Università dell’Aquila, via Vetoio, I-67100, L’Aquila, Italy}
\affiliation{INFN, Laboratori Nazionali del Gran Sasso, I-67100 Assergi (AQ), Italy}

\author{Violetta Sagun}\email{v.sagun@soton.ac.uk}
\affiliation{Mathematical Sciences and STAG Research Centre, University of Southampton, Southampton SO17 1BJ, United Kingdom}

\author{Kalin V. Staykov}\email{kstaykov@phys.uni-sofia.bg}
\affiliation{Department of Theoretical Physics, Sofia University ``St. Kliment Ohridski", Sofia 1164, Bulgaria}

\author{Daniela D. Doneva}\email{daniela.doneva@uni-tuebingen.de}
\affiliation{Departamento de Astronom\'ia y Astrof\'isica, Universitat de Val\`encia, Dr. Moliner 50, 46100, Burjassot (Val\`encia), Spain}
\affiliation{Theoretical Astrophysics, Eberhard Karls University of Tübingen, Tübingen 72076, Germany}

\author{Stoytcho S. Yazadjiev}\email{yazad@phys.uni-sofia.bg}
\affiliation{Department of Theoretical Physics, Sofia University ``St. Kliment Ohridski", Sofia 1164, Bulgaria}
\affiliation{Institute of Mathematics and Informatics, Bulgarian Academy of Sciences, Acad. G. Bonchev St. 8, Sofia 1113, Bulgaria}

\begin{abstract}
Dark matter is expected to accumulate inside neutron stars, modifying the structure of isolated stars and influencing both the dynamics of binary mergers and the evolution of the resulting hypermassive remnants. Since differential rotation is the primary mechanism delaying the collapse of these remnants, understanding its behavior is crucial when assessing the impact of an embedded dark component. In this work, we extend the numerical code {\tt RNS} to describe two gravitationally coupled fluids in differential rotation, with baryonic matter modeled by a realistic nuclear equation of state and dark matter represented as a self-interacting bosonic condensate. Within this framework, we construct equilibrium sequences for a representative differential rotation law, providing a basis to explore how dark matter may influence the global properties and rotational dynamics of binary neutron star remnants.
\end{abstract}

\maketitle

\section{Introduction\label{sec:intro}}
Neutron stars (NSs) represent one of the most extreme environments in the Universe. Their mergers are key laboratories for strong field gravity, dense matter physics, and multimessenger astrophysics. The detection of GW170817~\cite{LIGOScientific:2017vwq} confirmed binary neutron star (BNS) coalescence as a source of both gravitational waves (GWs) and electromagnetic counterparts, providing constraints on the dense matter equation of state (EOS)~\cite{Bauswein:2017vtn, Radice:2017lry}. Depending on the total mass and the underlying EOS, the post-merger remnant may form a hypermassive neutron star (HMNS), temporarily stabilized against collapse by differential rotation and thermal pressure~\cite{Shibata:1999wm, Baiotti:2016qnr, Paschalidis:2016vmz}, or promptly collapse into a black hole. Differential rotation is the dominant mechanism supporting such remnants, and its properties play a central role in determining their stability and lifetime. Uniformly rotating stars, on the other hand, are limited by the Keplerian frequency, the rotation rate at which matter at the equator follows a geodesic orbit, beyond which any further spin-up results in mass shedding.

On the other hand, differential rotation can sustain configurations well above this threshold~\cite{Baumgarte:1999cq, Paschalidis:2016vmz}, producing hypermassive stars that survive for tens to hundreds of milliseconds before collapsing under the combined effects of angular momentum redistribution, magnetic breaking, and GW emission. Early studies of differentially rotating stars often relied on the “$j$-constant” law, in which the specific angular momentum is uniform throughout the star, leading to an angular velocity that decreases approximately as the inverse square of the distance from the rotation axis~\cite{10.1093/mnras/239.1.153}. However, simulations of BNS mergers have shown that the resulting HMNS remnants exhibit a more complex rotational structure: a slowly rotating, nearly uniform core, an outer region with rapidly increasing angular velocity, and a radial decline consistent with a Keplerian profile~\cite{PhysRevD.91.064027, PhysRevD.96.043004, PhysRevD.95.063016, PhysRevLett.120.221101, PhysRevD.100.124042, PhysRevD.91.064001}. To capture these features more realistically, several parameterized prescriptions for differential rotation have been introduced~\cite{Uryu:2017, 10.1093/mnras/staa2725, PhysRevLett.124.171103,Cassing:2024dxp}, which reproduce the angular velocity profiles of merger remnants accurately for both polytropic and tabulated EOS~\cite{ PhysRevD.103.063014,10.1093/mnras/stab392, 10.1093/mnras/stab3565}.

On the other hand, many particle physics scenarios suggest that NSs may also host a dark matter (DM) component. In asymmetric DM models, particles accreted during a star’s lifetime do not annihilate and therefore accumulate steadily in the stellar interior~\cite{PhysRevD.89.015010, Bell:2020obw, Bell:2020jou, Stref:2016uzb, Lacroix:2018zmg}. The efficiency of this process depends both on environmental factors, such as the local DM density, and on microphysical inputs, such as scattering cross-sections, production during core-collapse supernovae, or capture in the progenitor star~\cite{Nelson_2019, PhysRevD.97.123007}. Self-interacting DM is particularly compelling, as it can simultaneously address small-scale tensions, e.g., the core-cusp problem in dwarf galaxies~\cite{DelPopolo:2021bom}, in the cold DM paradigm, and provide sufficient pressure to form stable structures and prevent collapse into a black hole~\cite{Suarez:2013iw}. Depending on the particle mass and self-interaction strength, the dark component may settle into a compact core embedded in the NS or extend into a diffuse halo~\cite{Hippert:2022snq, Liu:2024rix, Koehn:2024gal}. In either case, the presence of DM alters macroscopic observables such as the mass-radius relation and the moment of inertia, mimicking the effect of softening (in the case of the core morphology) and stiffening the baryonic EOS (in the case of the halo morphology)~\cite{Bezares:2019jcb, DiGiovanni:2022mkn, Ruter:2023uzc}.

Previous works have investigated static or uniformly rotating DM-admixed NSs, as well as their role during the inspiral phase of BNS mergers~\cite{Bezares:2019jcb, Bauswein:2020kor, Emma:2022xjs, Routaray:2024lni, Cronin:2023xzc, Konstantinou:2024ynd, Mourelle:2024qgo, Cipriani:2025tga, Giangrandi:2025rko}. The case of long-lived differentially rotating remnants remains largely unexplored. The combined effect of differential rotation with a dark component opens new dynamical possibilities. A compact DM core could deepen the gravitational potential and accelerate collapse, while an extended halo could modify the moment of inertia and the spectrum of the remnant post-merger oscillations. The presence of DM may also shift stability boundaries and alter the maximum mass that HMNS can reach.  Because the GW signal encodes the internal structure and composition of the remnant, these effects could leave detectable imprints on the kilohertz GW spectrum targeted by current and next-generation detectors~\cite{ET:2025xjr}. 

Although dynamical simulations are ultimately required to capture the nonlinear evolution of merger remnants, they are computationally expensive and restricted to limited sets of initial conditions. Quasi-equilibrium models offer a complementary approach to explore the parameter space systematically. They have already proven useful in interpreting post-merger GW spectra, estimating threshold masses for prompt collapse, and constructing empirical relations between remnants and nonrotating stars ~\cite{Paschalidis:2016vmz,Bozzola:2017qbu,Weih:2017mcw,10.1093/mnras/stab392,Ciolfi2021, Rosati2021}. Extending this methodology to two-fluid systems enables us to investigate how DM modifies these universal relations and to identify possible multimessenger signatures. Our study, therefore, occupies an intermediate ground between microphysical modeling and full merger simulations, providing a controlled framework to isolate the impact of DM of various fractions and rotation law on HMNS stability and structure.  

In this work, we construct equilibrium sequences of differentially rotating, DM-admixed NSs using an extended version~\cite{Cipriani:2025tga} of the \texttt{RNS} code~\cite{Stergioulas:1994ea, Stergioulas:2003ep}, adapted to treat two gravitationally coupled fluids. We focus on representative configurations that highlight the role of the dark component in shaping the maximum mass, angular velocity profile, and stability boundaries of HMNS remnants. We should note that this approach is different from the DM admixed NSs modeled as fermion-bosonic compact objects~\cite{Henriques:1989ez,Mourelle:2024qgo}. A study of differentially rotation in this case is also underway~\cite{Mourelle:2025}.

The paper is organized as follows. In Section~\ref{sec:setup} we describe our numerical setup and the assumptions adopted for the differential rotation law. In Sections~\ref{sec:BMEOS} and~\ref{sec:DMEOS} we present the EOSs for baryonic matter (BM) and DM, respectively. Section~\ref{sec:results} discusses the equilibrium sequences obtained and their astrophysical implications. Finally, Section~\ref{sec:conclusions} summarizes our findings and outlines future directions.

Unless otherwise stated, we employ geometrized units with $G = c = 1$ and express all quantities in solar-based units. Energy densities and pressures are shown in cgs units to facilitate comparison with the standard tabulated nuclear EOS. For the DM EOS, the natural units $\hbar = c = 1$ are used.

\section{Theoretical framework\label{sec:setup}}
Employing the Komatsu-Eriguchi-Hachisu (KEH) scheme~\cite{Komatsu:1989zz} with the modifications introduced in~\cite{Cook:1992}, the Einstein's field equations can be solved within stationary and axial symmetry. The generic line element takes the form
\begin{equation}
\begin{split}
    \dd s^2 & = g_{\mu\nu} \dd{x^\mu}\dd{x^\nu} \\
     &= - e^{\gamma + \rho}\dd t^2 + e^{2\alpha} (\dd r^2 + r^2 \dd \theta^2)\\
     &\hspace{11pt}+ e^{\gamma - \rho} r^2 \sin^2\theta(\dd \phi - \omega \dd t)^2,
\end{split}
\end{equation}
where $\gamma, \rho$, $\mu$ and $\omega$ are the metric fields which depend on $r$ and $\theta$. The metric fields and matter distribution are computed iteratively starting from an initial guess, typically a nonrotating star.

To model numerically differentially rotating DM admixed NSs, we expand on the work presented in~\cite{Cipriani:2025tga}, where the {\tt RNS} code was extended to model uniformly rotating NSs with DM. Denoting the specific enthalpy of the fluid $\X \in \{\BM,\,\DM\}$ as $H_\X$, the fluid's four-velocity as $u^\mu_\X$ and the specific angular momentum $j_\X = g_{\mu\phi} u^t_\X u^\mu_\X$, the first integral of the hydrostationary equilibrium is written as
\begin{equation}\label{eq:hydroEq}
    H_\X - \ln{u^t_\X} + \int_{0}^{j} \tilde{j}_\X \frac{\dd \Omega_\X}{\dd \tilde{j}_\X} \dd \tilde{j}_\X = \text{const}.\
\end{equation}
Coupling Eq.~\eqref{eq:hydroEq} with an explicit expression for the angular velocity $\Omega(j)$ and an EOS, an updated matter distribution can be found. Then, the metric potentials are recomputed. This procedure continues until convergence is reached. For a more comprehensive description of the algorithm, see for example~\cite{Komatsu:1989zz,10.1093/mnras/239.1.153, 10.1093/mnras/stab392, Cassing:2024dxp}.

Of utmost importance is the choice of the angular velocity profile $\Omega(j)$. Many laws have been proposed, aimed at describing the profiles found in proto-NSs and remnants of BNS mergers. The classic ``$j$-constant" law is
\begin{equation}
    j(\Omega) = A^2 (\Omega_c - \Omega)
\end{equation}
where $A$ is a positive constant that determines the length scale over which the angular velocity varies within the star and $\Omega_c$ is the angular velocity at the rotation axis. Although this law describes the proto-NS profile well~\cite{Villain:2003ey}, it falls short in the case of BNS mergers~\cite{10.1093/mnras/stab392, Iosif:2021-09312}. 

A particularly important family of laws has been proposed in~\cite{Uryu:2017}: they all feature a peak in the rotation profile and a Keplerian fall-off. We will employ the following rotation law:
\begin{equation}\label{eq:rotLaw}
    \Omega(j) = \Omega_c \frac{1 + \frac{j}{B^2 \Omega_c}}{1 + \left(\frac{j}{A^2 \Omega_c}\right)^4}
\end{equation}
that has been widely used in the literature.

Throughout this work, we employ dimensionless quantities to characterize the DM and BM components of the star. For the global properties, we define
\begin{align}
f^{M}_{\rm DM} &= \frac{M_{\rm DM}}{M_{\rm BM} + M_{\rm DM}}, \label{eq:fM}\\
f^{J}_{\rm DM} &= \frac{J_{\rm DM}}{J_{\rm BM} + J_{\rm DM}}, \label{eq:fJ}
\end{align}
which represent, respectively, the fraction of the total mass and the fraction of the total angular momentum carried by the DM component. These parameters provide a compact way to quantify the relative importance of DM in the global equilibrium, independent of the absolute values of the mass and angular momentum.

To characterize the rotational structure of each fluid, we introduce two additional dimensionless ratios,
\begin{align}
\lambda^{X}_{1} &= \frac{\Omega^{X}_{\max}}{\Omega^{X}_{c}}, \label{eq:lambda1}\\
\lambda^{X}_{2} &= \frac{\Omega^{X}_{e}}{\Omega^{X}_{c}}, \label{eq:lambda2}
\end{align}
with $X \in \{{\rm BM, DM}\}$. Here $\Omega^{X}_{c}$ is the angular velocity on the rotation axis, $\Omega^{X}_{e}$ the value at the equator, and $\Omega^{X}_{\max}$ the maximum angular velocity attained inside the star measured by an observer at infinity. The parameters $\lambda^{X}_{1}$ and $\lambda^{X}_{2}$, therefore, capture the degree of differential rotation of the fluid: $\lambda_{2}$ measures how rapidly the equatorial layers rotate relative to the core, while $\lambda_{1}$ determines the strength and location of the peak in $\Omega(r)$. In the one-fluid case, these two ratios are commonly used to classify differentially rotating equilibria into morphological families (Types A, B, C, and D)~\cite{Ansorg:2009, 2017ApJ...837...58G}, and in our two-fluid framework, they play the same role in identifying whether a configuration is quasi-spherical or toroidal. Our setup in \texttt{RNS} can recover only two of those. Type A solutions correspond to stars featuring a maximum of the energy density at the center. Unlike other solution types, their overall morphology remains quasi-spherical, but the internal distribution of angular velocity can vary substantially depending on the rotation parameters. On the other hand, Type C solutions exhibit toroidal-like structures with off-center density maxima, where stronger differential rotation significantly alters the star’s shape and allows for higher maximum masses. In this work, we construct sequences of both types, covering the diversity of morphologies permitted by differential rotation and systematically exploring how the presence of a DM component modifies their equilibrium and global properties.

\subsection{BM EOS}\label{sec:BMEOS} 
The properties of hadronic matter are modeled using the relativistic density functional DD2 EOS~\cite{PhysRevC.81.015803}. This EOS includes neutrons, protons, electrons, and muons and is known for its relatively stiff behavior at high densities. It satisfies the maximum mass constraints derived from observations of massive NSs~\cite{Antoniadis:2013pzd, Romani:2021xmb, Romani:2022jhd}, the limits set by GW measurements~\cite{LIGOScientific:2018cki, LIGOScientific:2020aai} and NICER observations~\cite{Miller:2019cac, Riley:2019yda, Miller:2021qha, Riley:2021pdl, Choudhury:2024xbk}. 

For densities below nuclear saturation, the DD2 EOS is supplemented by a crust model based on the generalized relativistic density functional framework~\cite{Typel:2018wmm}. This crust description features nuclei arranged in a body-centered cubic lattice immersed in a uniform electron background, with an additional neutron gas component appearing above the neutron drip density. The transition between the crust and the core EOSs is implemented within a unified, consistent framework.

These choices ensure a physically motivated modeling of BM, capturing the relevant nuclear physics from the crust to the core while respecting current astrophysical constraints.

\subsection{DM EOS}
\label{sec:DMEOS}

The DM component in this work is modeled as self-interacting bosons governed by the Lagrangian~\cite{Colpi:1986ye}
\begin{equation}\label{eq:DMLagr}
    \mathcal{L} = \frac{1}{2}\partial_\nu \phi^* \partial^\nu \phi - \frac{m_{\DM}^2}{2} \phi^* \phi - \frac{\lambda}{4} (\phi^* \phi)^2,
\end{equation}
where $\phi$ is a complex scalar field, $m_{\DM}$ is the DM particle mass, and $\lambda$ is a dimensionless coupling constant.

At sufficiently low temperatures, the scalar field forms a Bose-Einstein condensate, enabling stable configurations. A detailed derivation of the EOS in the zero-temperature (total condensation) limit is provided in~\cite{Karkevandi:2021ygv}. Introducing the chemical potential $\mu_{\DM}$, the key thermodynamic quantities are expressed as
\begin{subequations}\label{eq:DMEOSAll}
\begin{align}
    n_{\DM} &= \frac{\mu_{\DM}}{\lambda} \left(\mu_{\DM}^2 - m_{\DM}^2\right), \\
    P_{\DM} &= \frac{1}{4 \lambda} \left(\mu_{\DM}^2 - m_{\DM}^2\right)^2, \\
    \varepsilon_{\DM} &= 3 P_{\DM} + 2 m_{\DM}^2 \sqrt{\frac{P_{\DM}}{\lambda}}, \\
    h_{\DM} &= \log \frac{\mu_{\DM}}{m_{\DM}},
\end{align}
\end{subequations}
which correspond to the DM number density, pressure, energy density, and specific enthalpy, respectively.

\begin{figure*}
    \centering
    \includegraphics[width=1\linewidth]{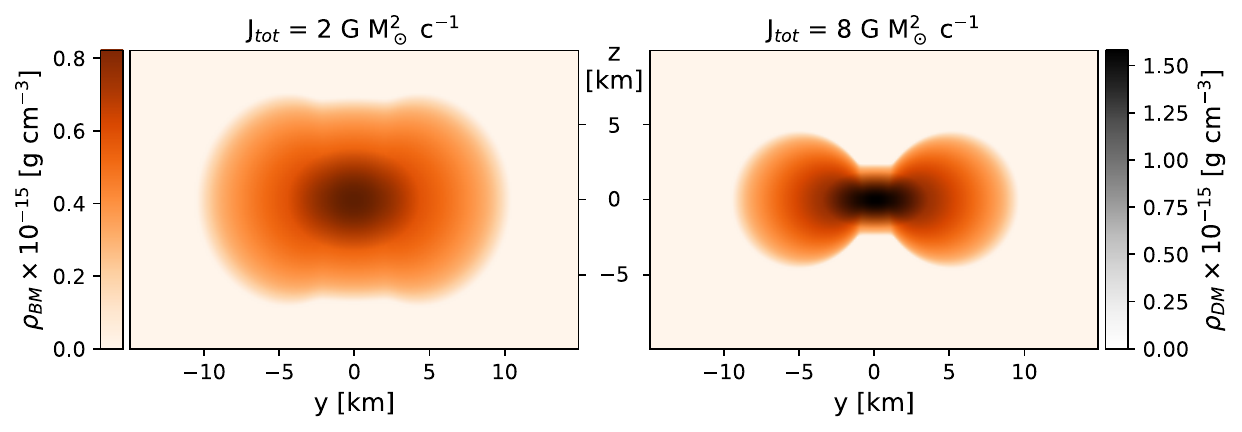}
    \caption{BM and DM density profiles in the \textit{yz} plane of two configurations with the same baryonic central energy density $\epsilon^c_\BM = \SI{0.75e15}{\dens}$ and $\fm = 5\%$, $\fj = 1\%$, $(\lambda_1, \lambda_2) = (2, 0.5)$.}
    \label{fig:profiles_(2,0.5)}
\end{figure*}

\section{Results\label{sec:results}}
Differentially rotating NSs described by the rotational law \eqref{eq:rotLaw} can serve as proxies for remnants formed in BNS mergers. Since we are interested in stationary equilibrium configurations, it is natural to compare with such merger remnants as obtained in numerical relativity simulations. A relevant example is provided in \cite{Giangrandi:2025rko}, which considers two-fluid systems in both core and halo configurations. Among the models discussed there, we focus on those featuring a DM core, as they exhibit a simple structure in which the DM component rotates almost uniformly, while the BM retains its characteristic differential rotation. This qualitative behavior is also visible in the bottom-right panel of Fig.~5 in~\cite{Giangrandi:2025rko}, which shows the instantaneous angular velocity profiles of both fluids in a dynamical simulation. Although in that case the DM core appears nearly rigidly rotating, its profile evolves in time, and the slope varies slightly throughout the evolution. We nonetheless adopt the assumption of uniform rotation. For the remainder of this work, we will therefore assume that Eqs.~\eqref{eq:lambda1} and \eqref{eq:lambda2} refer exclusively to the BM component, and we omit the corresponding tag to simplify the notation. We leave halo configurations for future work, as their angular velocity profiles cannot be reproduced with the rotation laws currently implemented in {\tt RNS}. These profiles feature a maximum displacement from the origin and an angular velocity that increases linearly near the center. In contrast, in BM configurations, the core rotates almost uniformly, with strong differential rotation confined to the outer layers.

\subsection{Toroidal configurations}
Figure~\ref{fig:profiles_(2,0.5)} shows two NS representative configurations with a DM core with different angular momenta -- one with slower rotation (left), keeping a more spherical-like shape, and one with a very high angular momentum (right), making the star highly elongated. The specific choice of $\lambda_1 = 2$ and $\lambda_2 = 0.5$ gives rise, for sufficiently high angular momentum, to the characteristic toroidal shape, commonly referred to as \textit{Type C solution} (see, e.g., \cite{Ansorg:2009, 10.1093/mnras/stab3565}) where an off-center maximum of the energy density is observed, as shown in Fig.~\ref{fig:profiles_(2,0.5)}. The angular velocity profile is characterized by an almost uniformly rotating core, followed by a sharp rise to a maximum value determined by $\lambda_1$, and then a slow decrease approaching the Keplerian law $\Omega \propto r^{-3/2}$. In both panels of Fig.~\ref{fig:profiles_(2,0.5)}, the DM component carries 1\% of the total angular momentum $J_\tot$ and maintains a quasi-spherical shape, as it rotates uniformly apart from small deviations induced by frame dragging.

Figure~\ref{fig:Me_DD2noY_(2,05)} shows equilibrium sequences for selected configurations in the $(M, \epsilon^c_\BM)$ plane at various values of the total angular momentum $J_\tot$, ranging from 0 to $\SI{8}{\Junit}$, each identified by one color sampled from black to aquamarine. Dashed lines represent, for reference, the one-fluid description of isolated NSs; red dots indicate the maximum mass for each sequence. Solid (dash-dotted) lines correspond to models with constant values of the DM to BM ratio $\fm = 5\%$, while the angular moment fraction is $\fj = 1\%$ ($\fj = 5\%$). Red circles (triangles) on each line mark their respective maximum mass. Finally, dotted lines illustrate configurations with constant $\fm = 5\%$ and much higher, near Keplerian limit, rotation of the DM component $\Omega_\DM = 0.75\, \Omega_\K^\DM$; their maximum masses are indicated by stars. 

\begin{figure}
    \centering
    \includegraphics[width=1\linewidth]{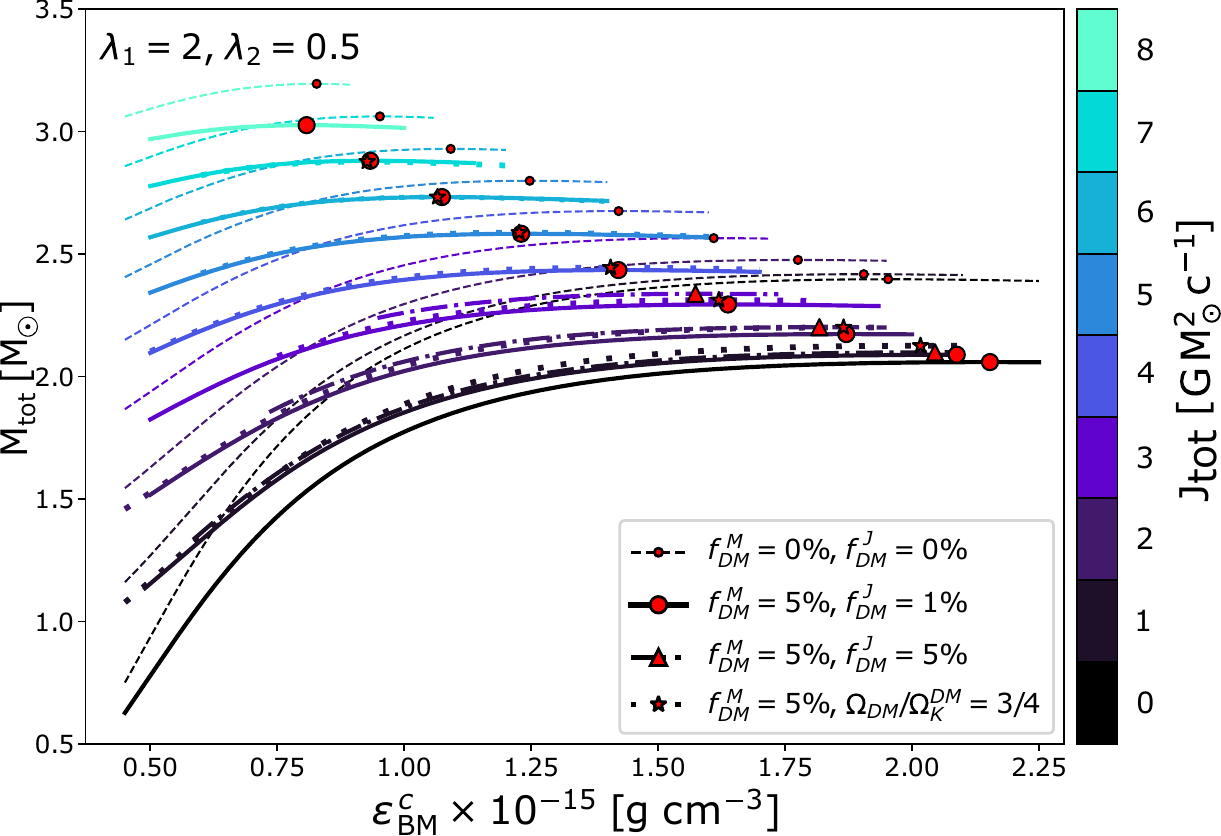}
    \caption{Mass vs BM central energy density sequences for the DD2 EOS and rotational parameters $\lambda_1 = 2$ and $\lambda_2 = 0.5$. The sequences have fixed total angular momentum $J_\tot$, represented with the color gradient. Dashed lines represent solutions with $\fm = 0\%$, solid lines show solutions with $\fm = 5\%$ and $\fj = 1\%$. Both have been computed up to $J_\tot = \SI{8}{\Junit}$. Dash-dotted lines represent solutions with $\fm = 5\%$ and $\fj = 5\%$, computed up to $J_\tot = \SI{4}{\Junit}$ and finally dotted lines solutions with $\fm = 5\%$ and $\Omega_\DM / \Omega^\DM_\K = 3/4$, computed up to $J_\tot = \SI{7}{\Junit}$. The red symbols report the sequence maximum mass, marking the turning point beyond which models become unstable. Since a nonlinear monotonic relationship exists between central and maximum energy density for all reported models, we present the results as a function of the former, which is the parameter used to generate the models.}
    \label{fig:Me_DD2noY_(2,05)}
\end{figure}

\begin{figure}
    \centering
    \includegraphics[width=1\linewidth]{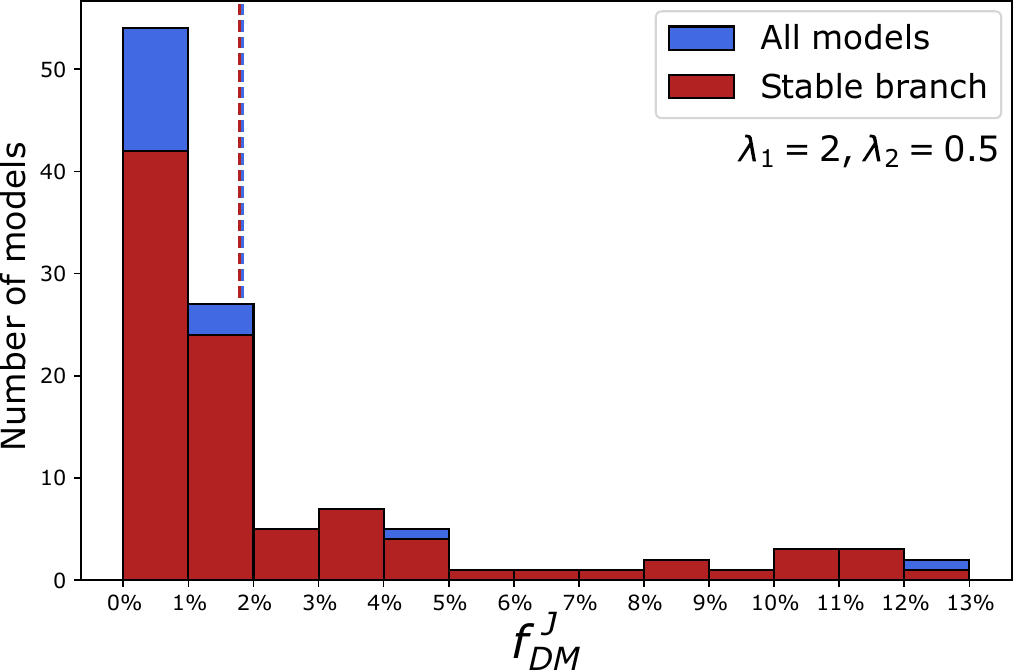}
    \caption{Distribution of the angular momentum fraction in the DM component for models rotating at $75\%$ of their Kepler frequency. Blue and red bars represent the full sample and the stable models only, respectively. Dashed lines indicate the average values: $\fj = 2.68\%$ for the former and $\fj = 2.65\%$ for the latter. Models with $\fj > 4\%$ exhibit low total angular momentum ($J_\tot < \SI{2}{\Junit}$) and lie near the maximum of their respective sequence. These models are sampled uniformly along the dotted sequences shown in Fig.~\ref{fig:Me_DD2noY_(2,05)}.}
    \label{fig:histogram}
\end{figure}

We adopt the DM mass fraction of $\fm = 5\%$, consistent with the values used in~\citet{Karkevandi:2021ygv,Giangrandi:2025rko}, in order to enable a meaningful comparison. This value is consistent with observations of $\SI{2}{\Msun}$ NSs together with $\Lambda_{1.4} \le 580$ constraint~\cite{LIGOScientific:2018cki}, set by LIGO-Virgo Collaboration. For the angular momentum partition, we adopt a fiducial $\fj = 1\%$ to probe the impact of a slowly rotating, quasi-uniform DM component, and we also show sequences with $\fj = 5\%$ to illustrate the trend when the DM carries a larger fraction of the total rotation. The choice $\fj=1\%$ is intentionally modest: with $\fm=5\%$ the DM core cannot (within our microphysical assumptions) sustain arbitrarily large angular momentum. Setting $\fj = 5\%$ already eliminates most of the previously accessible parameter space, as can be seen looking at Fig.~\ref{fig:Me_DD2noY_(2,05)}.

Finally, the dotted sequences with $\Omega_\DM=0.75\,\Omega_\K^\DM$ represent an extreme, near-Kepler case for a rapidly rotating DM component. Figure~\ref{fig:histogram} shows the distribution of these models as a function of $\fj$. The blue histogram includes all models from Fig.~\ref{fig:Me_DD2noY_(2,05)}, while the red bars indicate only those on the stable branches. Most models have $\fj < 2\%$, with a few outliers reaching up to $\fj = 12.3\%$. These extreme cases occur exclusively in slowly rotating sequences, where even big fractions correspond to relatively small angular momenta. Among the computed models, only in sequences with $J_\tot < \SI{2}{\Junit}$ we obtain configurations near the maximum with $\fj > 4\%$. Specifically, for $J_\tot = \SI{1}{\Junit}$ this occurs for $M_\tot > \SI{1.64}{\Msun}$, and for $J_\tot = \SI{2}{\Junit}$ for $M_\tot > \SI{2.14}{\Msun}$, to be compared with the respective maximum stable masses of $\SI{2.12}{\Msun}$ and $\SI{2.19}{\Msun}$.

Within the explored parameter space, Fig.~\ref{fig:histogram} indicates that sequences with fixed $\fj = 1\%$ are representative of the typical behavior found in our models. Alternative assumptions, e.g. larger $\fm$, different DM rotation laws, or strong DM self-interaction, will only change the results quantitatively.

The inclusion of a DM component generally leads to a reduction in the maximum gravitational mass supported by the configuration. This is expected in the case of core-type distributions, where the additional gravitational pull from a centrally concentrated DM core effectively softens the system~\cite{RafieiKarkevandi:2021hcc}. The BM component experiences a stronger total gravitational potential without receiving a corresponding increase in central pressure support, leading to an earlier onset of instability.

With the increase of the total angular momentum $J_\tot$, the difference between the DM admixed and pure BM NS decreases if one compares sequences with the same $J_\tot$. The dependence of the maximum total gravitational mass on the total angular momentum is represented in Fig.~\ref{fig:MJ2_DD2noY_(2,05)}, that shows the normalized maximum mass $\hat M = M_{\max} / M_{\max, J=0}$ as a function of the squared normalized total angular momentum $\hat J^2_\tot = J^2_\tot / M_{\max, J=0}^4$. The different markers denote configurations with varying mass and angular momentum fractions in the DM component, as indicated in the legend.

The solid lines are fitted on the points using a [2/1] Pad\'e resum of the Hartle-Thorne expansion in $\hat J_\tot^2$~\cite{Friedman:2013xza}:
\begin{equation}\label{eq:fit}
    \hat M(\hat J_\tot^2) = 1 + \frac{\alpha \hat J_\tot^2 + \beta \hat J_\tot^4}{1 + \gamma \hat J_\tot^2},
\end{equation}
which captures the leading-order rotational corrections and ensures a controlled behavior at large $J_\tot^2$. The fitting parameters are reported in Table~\ref{tab:fitParams}.

\begin{table}
  \centering
  \begin{tabular}{ccccc}
    \toprule
    {$(\fm,\,\fj)\%$} & {$\alpha$} & {$\beta$} & {$\gamma$} & {$M_{\max, J=0}$ [\si{\solarmass}]} \\
    \midrule
    $(0,\,0)\%$ & 0.284 & 0.027 & 0.496 & 2.398 \\
    $(5,\,1)\%$ & 0.269 & 0.022 & 0.454 & 2.059 \\
    $(5,\,5)\%$ & 0.365 & 0.168 & 1.316 & 2.059 \\
    \midrule
    \multicolumn{1}{c}{$(5\%,\ \frac{\Omega_\DM}{\Omega^\DM_K} = 3/4)$} & 0.396 & 0.164 & 1.739 & 2.059 \\
    \bottomrule
  \end{tabular}
  \caption{Fitting parameters for the model. The constants $\alpha$, $\beta$, and $\gamma$ are reported in dimensionless units.}
  \label{tab:fitParams}
\end{table}

In the bottom panel, we plot the logarithm of the absolute value of residuals; the fits reproduce the numerical data with high accuracy, typically within $10^{-4}$. The exception is the sequence at constant fraction $\Omega_\DM / \Omega_\K^\DM$ (in red), which shows slightly larger deviations but still remains within acceptable bounds.

Overall, rotation softens the DM induced reduction of the maximum mass due to the DM component, and all configurations at constant fraction of angular momentum show a common scaling with $J^2_\tot$, albeit with subtle differences that depend on the DM rotational properties: small values of $\fj$ ($\sim\,1\%$, orange line) tend to slow down the increase of $M_\max$ with respect to the one-fluid case, while higher values ($\sim\,5\%$, green line) increase it. Curves at constant angular velocity (red line) show instead a completely different behavior, in which Eq.~\eqref{eq:fit} has a positive second derivative and a much steeper increase.

\begin{figure}
    \centering
    \includegraphics[width=1\linewidth]{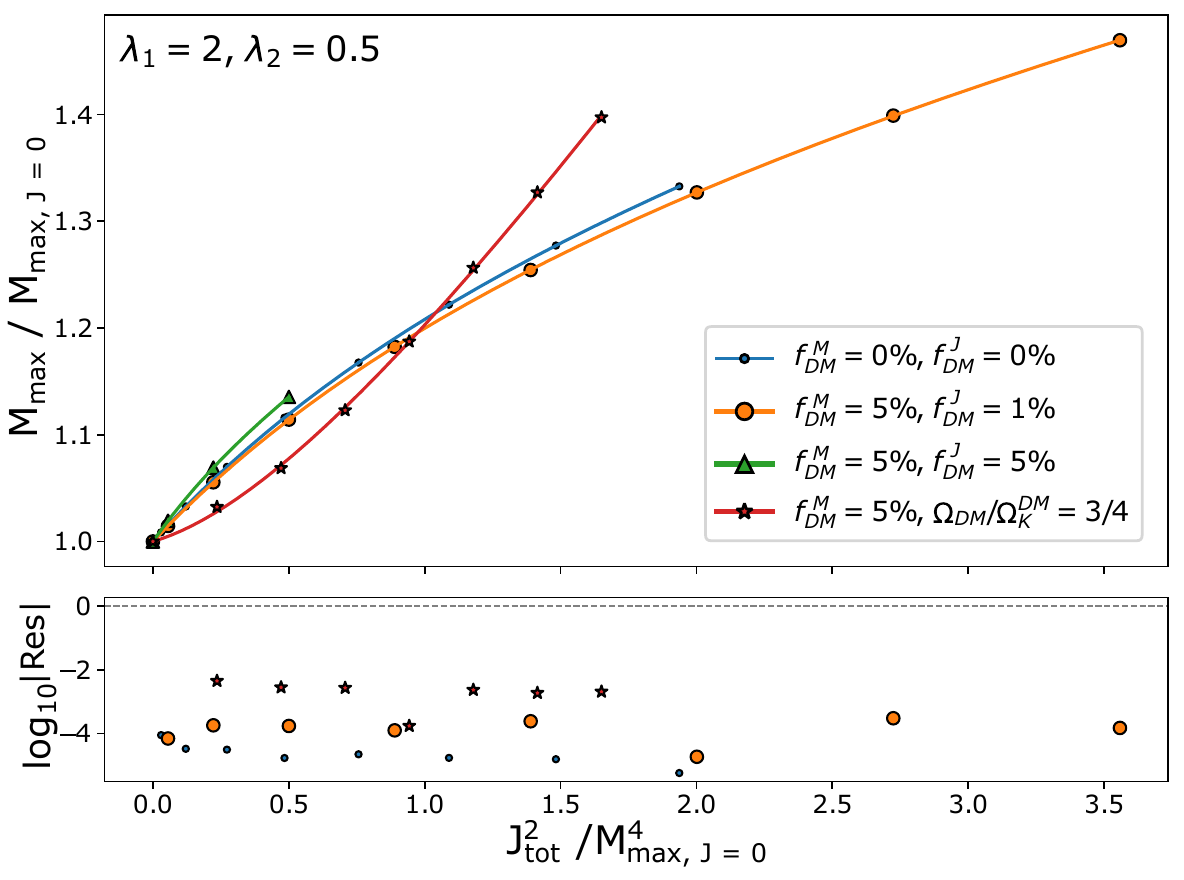}
    \caption{Normalized gravitational mass vs square of the total angular momentum for the same configurations computed in Fig.~\ref{fig:Me_DD2noY_(2,05)}. Colored lines are fitted with Eq.~\eqref{eq:fit}. The bottom panel shows the logarithm of the absolute value of the residuals.}
    \label{fig:MJ2_DD2noY_(2,05)}
\end{figure}

\begin{figure}
    \centering
    \includegraphics[width=1\linewidth]{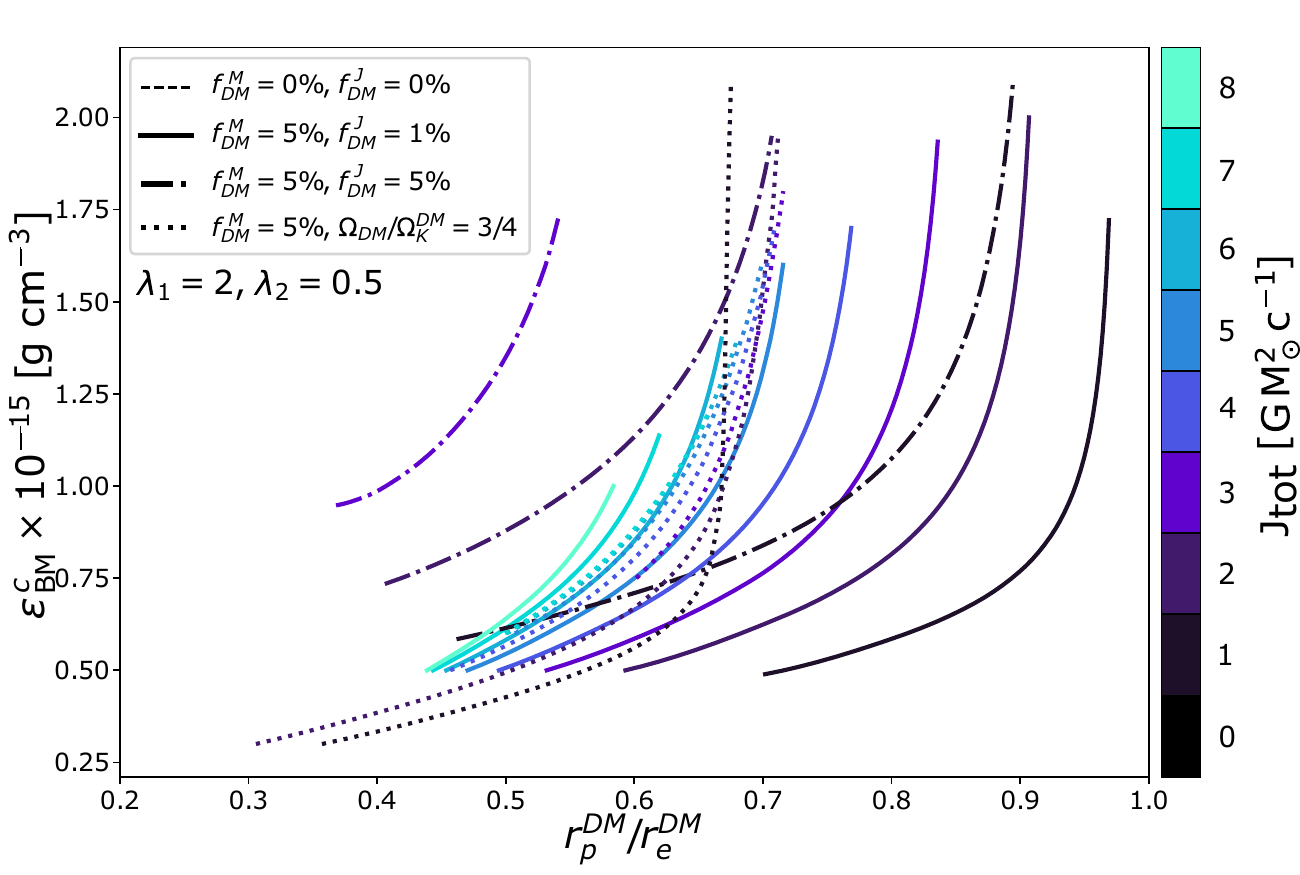}
    \caption{Central energy density of the BM component as a function of the DM deformation parameter. Lines and colors have the same meaning as in Fig.~\ref{fig:Me_DD2noY_(2,05)}.}
    \label{fig:eVr_DD2noY_(2,05)}
\end{figure}

\begin{figure*}
    \centering
    \includegraphics[width=1\linewidth]{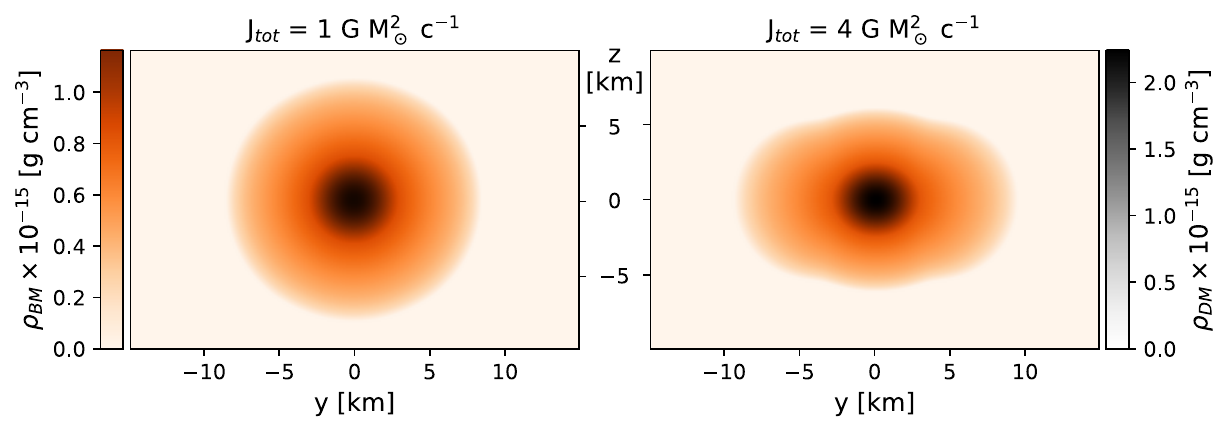}
    \caption{Matter density profiles in the \textit{yz} plane of two configurations with the same baryonic central energy density $\epsilon^c_\BM = \SI{1.47e15}{\dens}$ and $\fm = 5\%$, $\fj = 1\%$, $(\lambda_1, \lambda_2) = (1.8, 1)$.}
    \label{fig:profiles_(1.8,1)}
\end{figure*}

We now turn our attention to the sequences characterized by $\fj = 5\%$, shown in Fig.~\ref{fig:Me_DD2noY_(2,05)} (dash-dotted lines). As can be seen, we were able to compute sequences only for relatively low values of $J_\tot$, ranging from 0 to $\SI{3}{\Junit}$, and generally for a narrower range of central energy densities $\epsilon^c_\BM$. Figure~\ref{fig:eVr_DD2noY_(2,05)} provides further insight into the situation. It shows $\epsilon^c_\BM$ as a function of the resulting $\rrDM$, which reflects the deformation of the DM fluid. Increasing the target total angular momentum (and thus $J_\DM = \fj J_\tot$) shifts the sequence toward lower values of this ratio.

As previously observed in~\cite{Cipriani:2025tga}, the {\tt RNS} code faces difficulties when computing solutions involving very large deformations, even if the angular velocity remains below the Kepler limit. For smaller values of $\fj$, the code cannot compute configurations with $\epsilon^c_\BM < \SI{0.5e15}{\dens}$ despite their potentially large total masses, because building these configurations would require deformations beyond the code’s capability. Similarly, attempts to construct sequences with $J_\tot \geq \SI{4}{\Junit}$ at $\fj = 5\%$ lead to numerical problems with solutions having $\rrDM \lesssim 0.4$, a regime close to the apparent limit of the code’s numerical stability and convergence. Consequently, constructing DMANS with higher angular momentum is not possible for large $\fj$ within the current framework. This prevents us from reaching the realistically high angular momenta ($J > 5$) expected after a merger. This suggests that alternative differential rotation laws may be needed to plausibly model these high-$J$ scenarios.

\subsection{Quasi-spherical configurations}
In cases where the remnant of a BNS merger does not undergo immediate gravitational collapse to a black hole, it can form a hypermassive or supramassive NS that is sustained against gravitational collapse due to the rapid differential rotation. Due to the longer lifetime of such merger remnants, they have enough time to settle into a more quasi-spherical state before any delayed instability arises. Such configurations are often modeled using equilibrium sequences, particularly those classified as Type~A, which describe rotating stars retaining a maximum energy density at the center and remaining close to spherical symmetry despite increasing rotational flattening (quantified by $\rrBM$). To obtain these solutions, we adjust the parameters in \eqref{eq:lambda1} and \eqref{eq:lambda2} from $(2, 0.5)$ to $(\lambda_1, \lambda_2) = (1.8, 1)$. We choose these parameters following~\cite{PhysRevD.96.043004, 2020PhRvD.101f4052D, 10.1093/mnras/stab3565}, where they report that numerical simulations suggest that $\lambda_2 \approx 1$ and $\lambda_1 \in \left[ 1.7,\, 1.9\right]$ better approximate remnants from BNS merger. Two models are shown in Fig.~\ref{fig:profiles_(1.8,1)}.

Figure~\ref{fig:Me_DD2noY_(1.8,1)} shows sequences in the $(M_\tot, \epsilon^c_\BM)$ plane at fixed total angular momentum, with colors and line styles consistent with Fig.~\ref{fig:Me_DD2noY_(2,05)}. Maximum-mass points are marked with filled symbols and a black edge. For the quasi-spherical configurations corresponding to $(\lambda_1, \lambda_2) = (1.8, 1)$, we find that for $J_\tot < \SI{2}{\Junit}$ the resulting equilibrium sequences essentially coincide with those of the toroidal case shown in Fig.~\ref{fig:Me_DD2noY_(2,05)}. Although the angular-velocity profiles differ substantially, the global quantities such as $M_\tot$ remain nearly unchanged at low angular momentum, indicating that the detailed form of $\Omega(r)$ has only a minor influence in this regime.

\begin{figure}
    \centering
    \includegraphics[width=1\linewidth]{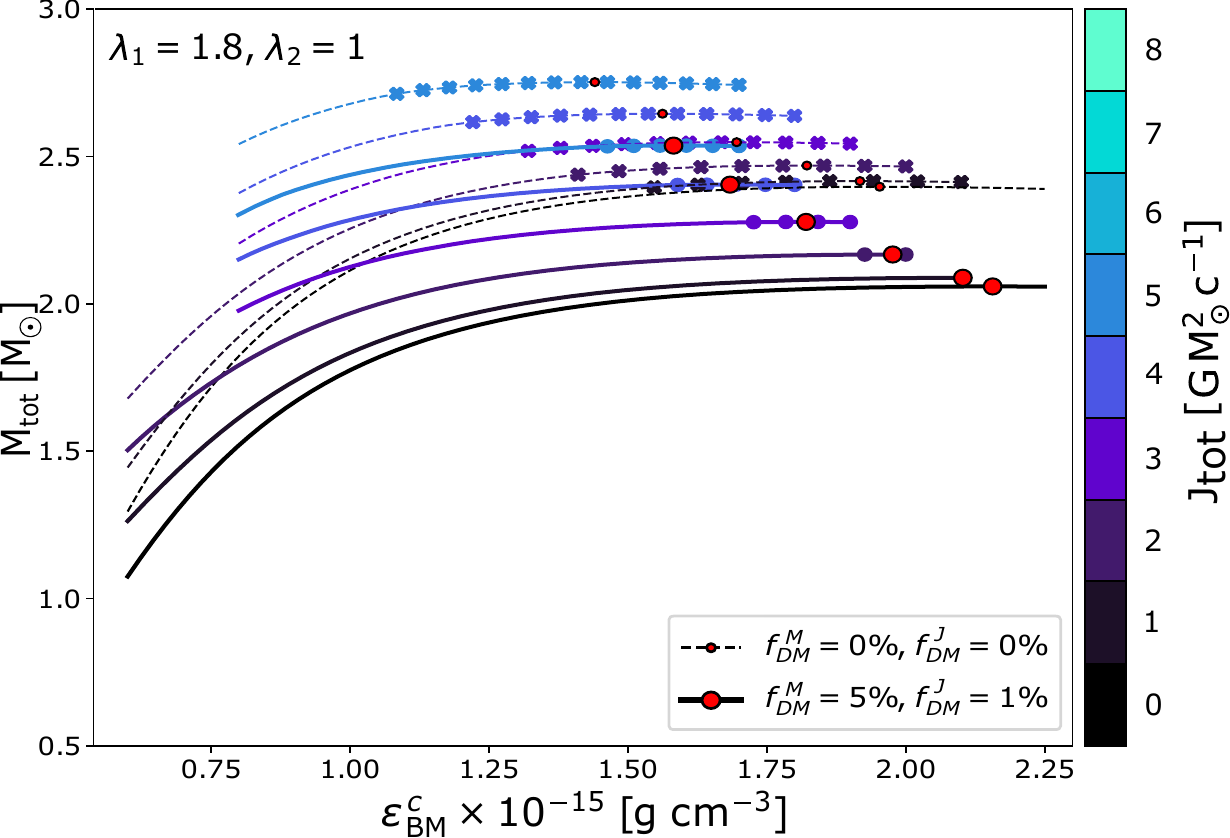}
    \caption{Gravitational mass vs baryonic central energy density for $\lambda_1 = 1.8$ and $\lambda_2 = 1$ at fixed total angular momentum $J_\tot$, represented with the color gradient. Dashed lines correspond to solutions with $\fm = 0\%$, solid lines to $\fm = 5\%$ and $\fj = 1\%$. Crosses and circles without the black edge represent cases for which the profile $\Omega(r)$ is discontinuous (\textit{cf}. Appendix~\ref{app:Error} and Fig.~\ref{fig:omega_of_r_profiles}) for $\fm = 0\%$ and $\fm = 5\%$, respectively.}
    \label{fig:Me_DD2noY_(1.8,1)}
\end{figure}

At higher values of $J_\tot$, up to the maximum accessible value of $\SI{5}{\Junit}$ in our numerical sequences, the maximum supported mass is consistently lower than in the Type~C case. This confirms, in the two-fluid scenario, the result already reported in~\cite{10.1093/mnras/stab3565, 2017ApJ...837...58G} for one-fluid stars: Type~A configurations are less efficient in exploiting differential rotation to increase the maximum mass.

\begin{figure*}[!t]
    \centering
    \includegraphics[width=1\linewidth]{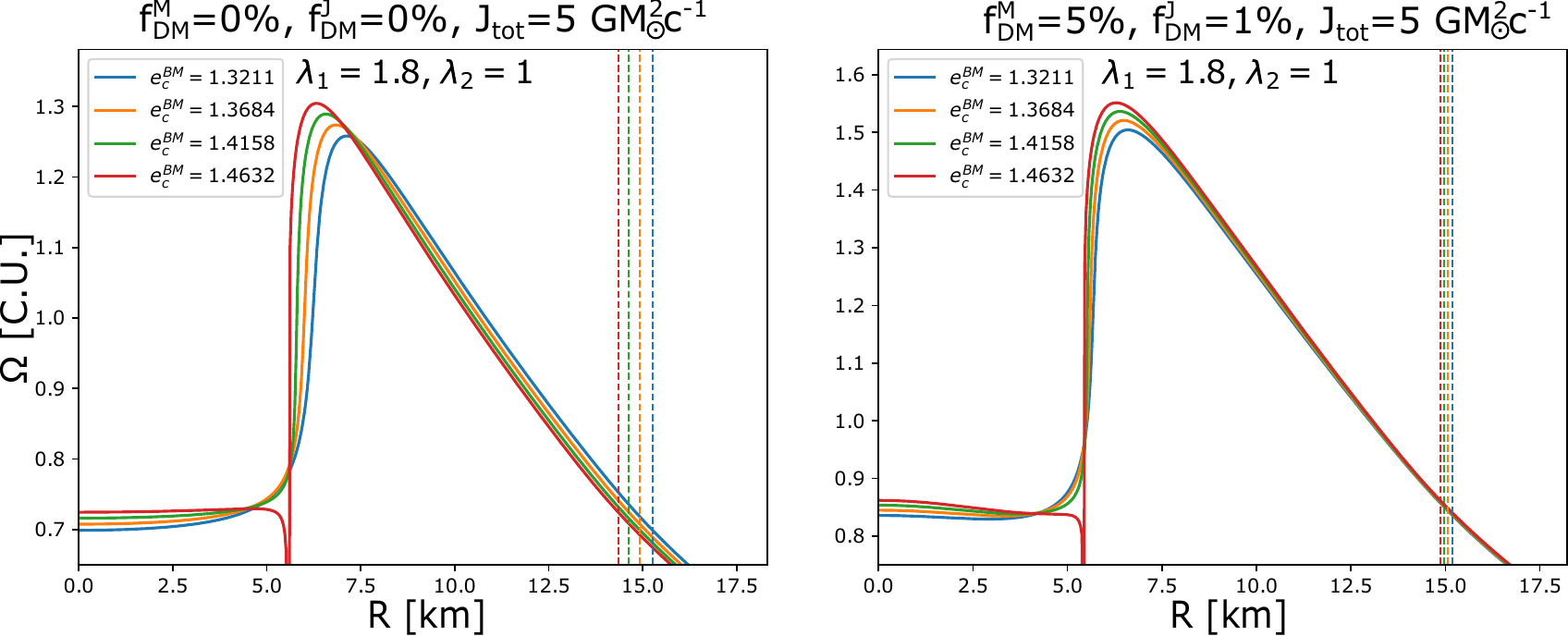}
    \caption{Angular velocity on the equatorial plane as a function of the proper radial position. Dashed lines mark the position of the star's surface. Left: example models extracted from Fig.~\ref{fig:Me_DD2noY_(1.8,1)} in the absence of DM. Right: same as the previous panel, but for a DM fraction of 5\%.}
    \label{fig:omega_of_r_profiles}
\end{figure*}

In the two-fluid case, even at moderate central densities, the BM equatorial angular-velocity profile $\Omega(r)$ develops a local minimum around $R\simeq\SI{4}{\kilo\meter}$ before rising sharply to its maximum and then approaching the Keplerian fall-off. This minimum reflects the influence of the DM component and is not produced by applying the rotation law of Eq.~\eqref{eq:rotLaw} to a single fluid, where $\Omega(r)$ instead increases monotonically outside the nearly flat core. Although similar minima can arise in one-fluid models under different rotation prescriptions~\cite{Cassing:2024dxp}, they do not appear under the assumptions adopted here. Figure~\ref{fig:omega_of_r_profiles} compares the equatorial profiles for the one-fluid (left) and two-fluid (right) descriptions across several values of the BM central energy density.

The minimum occurs because the DM slightly reduces the enclosed mass in the region where it is present, allowing $\Omega(r)$ to decrease locally while still satisfying the fixed total angular momentum $J_\tot$. Farther out, where $\Omega(r)$ reaches its maximum, the DM instead deepens the gravitational potential, requiring a steeper rise in $\Omega(r)$ for centrifugal support and producing a more pronounced peak than in the one-fluid case. This interplay between reduced and enhanced gravitational pull is visible in Fig.~\ref{fig:omega_of_r_profiles}.

Note that not all models along these sequences converge to physically valid solutions. In particular, for certain combinations of high central energy density and angular momentum, the rotation law in Eq.~\eqref{eq:rotLaw} ceases to admit real solutions for $\Omega(r)$ over a finite radial range. As a result, the corresponding equilibrium profiles develop discontinuities in the angular velocity, which we mark with open symbols in Fig.~\ref{fig:Me_DD2noY_(1.8,1)}. A discussion of the origin and impact of these discontinuities, together with a validation of our implementation against published equilibrium models, is presented in Appendix~\ref{app:Error}.

\section{Conclusions\label{sec:conclusions}}
In this work, we have studied DM-admixed NSs in the regime of differential rotation, which is of particular relevance for post-merger remnants of BNS coalescences. Building upon our earlier extension of the \texttt{RNS} code to two fluids, we have now implemented differential rotation laws that capture the angular velocity distributions observed in numerical relativity simulations of hypermassive NSs. In our framework, BM and DM are treated as two cold, gravitationally coupled fluids, with the baryonic component following differential rotation and the DM rotating quasi-uniformly, consistent with expectations from core-type configurations.

We considered sequences with fixed DM mass fractions up to 5\%, focusing on two representative choices of the rotation law parameters: $(\lambda_1, \lambda_2) = (2,0.5)$, which admits toroidal (Type C) configurations, and $(1.8,1)$, which yields quasi-spherical (Type A) equilibria. In both cases, the BM angular velocity reaches maximum away from the center, while the DM part rotates uniformly in accordance with numerical relativity merger simulations. Note that we focus only on the core DM configuration, i.e., when the DM component is roughly speaking confined inside the star. The reason is that halo configurations lead to a more complicated post-merger angular velocity distribution~\cite{Giangrandi:2025rko}. 

We investigated how the inclusion of a dark component modifies the maximum mass, stability, and structural properties of the remnant. Our results confirm that a centrally concentrated DM core generally reduces the maximum mass, as the additional gravitational pull is not compensated by extra pressure support. However, rotation tends to mitigate this reduction, and for sufficiently high total angular momentum the maximum masses of admixed stars approach those of their pure baryonic counterparts.

The scaling of the maximum mass with total angular momentum can be captured with high accuracy by a Padé resummation of the Hartle-Thorne expansion. This indicates that, despite the increased complexity of the two-fluid system, the global impact of a differential rotation remains predictable in terms of a small set of effective parameters. At the same time, the details of the angular velocity profile, and in particular the interplay between the two fluids, introduce new qualitative effects: we observe the emergence of a local minimum in the baryonic angular velocity profile, a feature absent in the corresponding one-fluid models. This effect arises from the redistribution of the gravitational potential by the dark component.

The limitations of the present analysis are primarily numerical: configurations with large DM angular momentum fractions or extreme deformations remain challenging to construct within the current {\tt RNS} framework. Nevertheless, the results confirm that the combination of differential rotation and a dark component produces imprints in equilibrium sequences and stability boundaries, with potential consequences for the interpretation of post-merger GW signals. However, equilibrium sequences alone are not sufficient to disentangle DM admixture from a simple softening of the baryonic EOS, since both scenarios lead to similar modifications in the mass-radius-moment of inertia plane.

Finally, we find that for large angular-momentum fractions, the central energy density, the numerical scheme approaches its limits; in particular, some quasi-spherical models exhibit discontinuities in $\Omega(r)$ where the adopted rotation law ceases to admit a real solution. Perhaps this is a residual of the employed differential rotation law and theory parameter. Thus, further investigation is needed, which is out of the scope of the present paper.

Future work can expand in several directions. First, the inclusion of halo-type configurations, which require more general rotation laws than currently available, would allow a broader survey of possibilities. Second, the dependence of our results on the DM microphysics, i.e., particle mass and self-interaction strength, remains to be mapped in detail, with the prospect of constraining regions of the parameter space through astrophysical observations. Finally, the equilibrium sequences constructed here can serve as initial data for full numerical relativity simulations of DM-admixed merger remnants. 

The equilibrium sequences studies provide a link between microphysical modeling and astrophysical observables in multimessenger astronomy. They allow for a systematic exploration of a very large parameter space covering DM properties, baryonic equations of state, and rotation laws, at a fraction of the computational cost of full dynamical simulations. The description of post-merger remnants as quasi-equilibrium models has already proven valuable in several contexts: interpreting the post-merger GW spectrum, determining the threshold mass to prompt collapse, constructing empirical relations that connect remnant properties to those of nonrotating models, and modeling longer-timescale processes relevant for multimessenger follow-up of GW detections~\cite{Paschalidis:2016vmz,Bozzola:2017qbu,Weih:2017mcw,10.1093/mnras/stab392,Ciolfi2021, Rosati2021}. In this sense, the equilibrium sequences developed here provide theoretical insight into the role of DM and represent a practical framework for connecting microscopic physics to future multimessenger observations.

\acknowledgments
This study is partly financed by the European Union-NextGenerationEU, through the National Recovery and Resilience Plan of the Republic of Bulgaria, project No. BG-RRP-2.004-0008-C01. DD acknowledges financial support via an Emmy Noether Research Group funded by the German Research Foundation (DFG) under grant no. DO 1771/1-1, by the Spanish Ministry of Science and Innovation via the Ram\'on y Cajal programme (grant RYC2023-042559-I), funded by MCIN/AEI/ 10.13039/501100011033, and by the Spanish Agencia Estatal de Investigaci\'on (grant PID2024-159689NB-C21) funded by the Ministerio de Ciencia, Innovaci\'on y Universidades. V.S. gratefully acknowledges support from the UKRI-funded ``The next-generation gravitational-wave observatory network'' project (Grant No. ST/Y004248/1). V.S. acknowledges the gratitude to the Funda\c c\~ao para a Ci\^encia e Tecnologia (FCT) I.P. for support under Advanced Computing Projects 2024.07037.CPCA.A1 with DOI identifier 10.54499/2024.07037.CPCA.A1 and 2023.10526.CPCA.A2 with DOI identifier 10.54499/2023.10526.CPCA.A2.

\appendix
\section{Analysis of angular velocity discontinuities}\label{app:Error}

Quasi-spherical models with $(\lambda_1, \lambda_2) = (1.8,\,1)$ lead to convergence issues even in the one-fluid scenario, especially at high central densities and angular momenta. 

Points marked with dots or crosses without black edges on the sequences (respectively for the two-fluid and one-fluid cases) of Fig.~\ref{fig:Me_DD2noY_(1.8,1)} correspond to models where the angular velocity profile $\Omega(r)$ becomes discontinuous, indicating failure of the iteration procedure. These discontinuities arise because the rotation law \eqref{eq:rotLaw} ceases to admit positive real solutions for $\Omega$ over a finite radial range, as illustrated in Fig.~\ref{fig:omega_of_r_profiles}, where we show different equatorial angular velocity profiles $\Omega$ as a function of the proper radius $R$ for a fixed total angular momentum $J = \SI{5}{\Junit}$. Colors correspond to different central BM energy densities, used as the sequence parameter. The left panel shows the one-fluid case ($\fm = \fj = 0$), and the right panel shows a two-fluid configuration with $\fm = 5\%$ and $\fj = 1\%$. 

\begin{figure}
    \centering
    \includegraphics[width=1\linewidth]{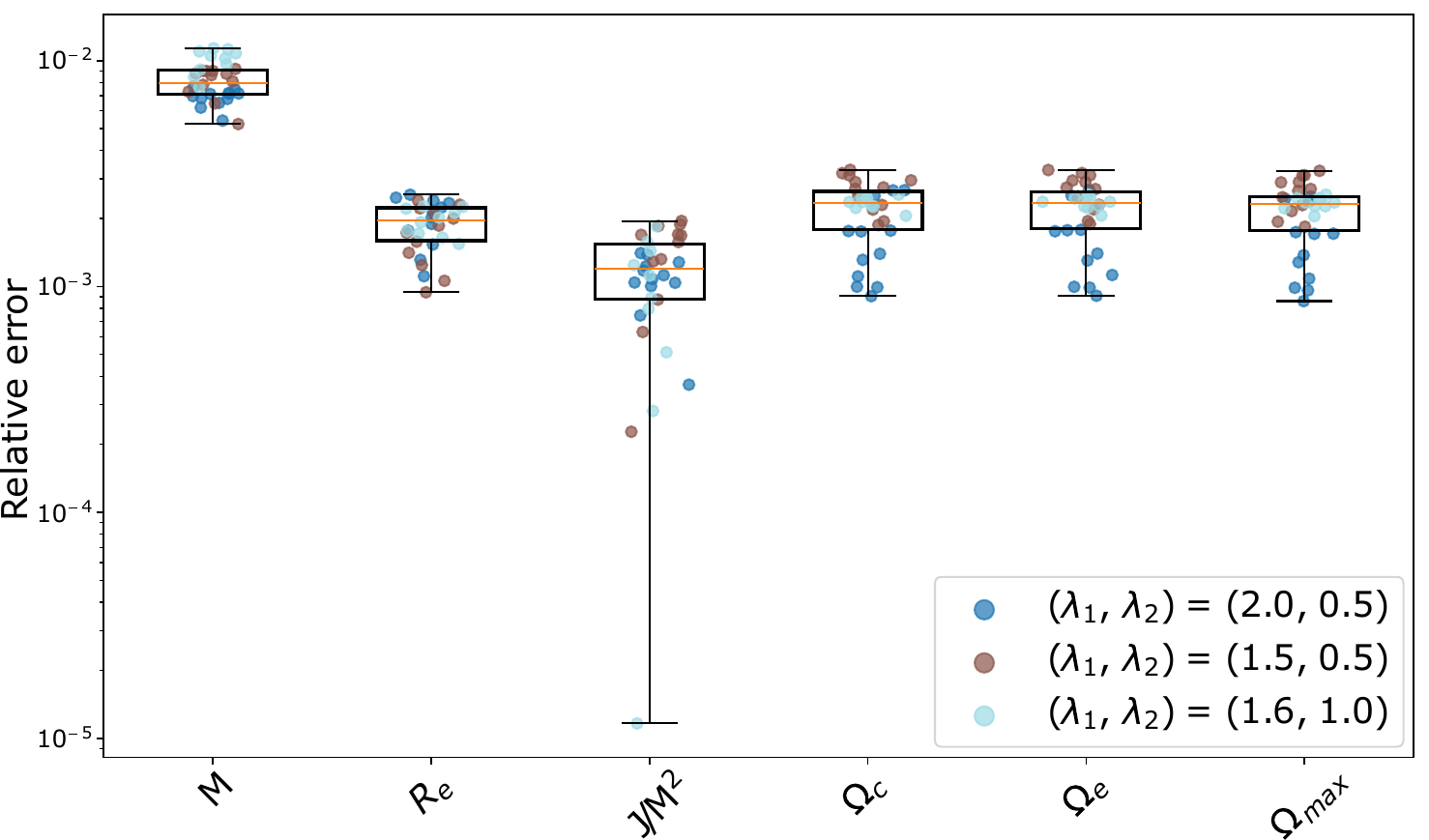}
    \caption{Relative errors of models computed in the one-fluid picture, compared against Table~6 of~\cite{10.1093/mnras/stab3565} for the DD2 EOS. We assess both global quantities (gravitational mass $M$ and dimensionless angular momentum $J/M^2$) and local ones (equatorial radius $R_e$, central and equatorial angular velocities, and the maximum angular velocity). Dots represent individual models, with each color corresponding to one of the three $(\lambda_1,\,\lambda_2)$ pairs considered. The orange line marks the median relative error, while the lower and upper edges of the box indicate the 25\% and 75\%, respectively.}
    \label{fig:Iosif_errors}
\end{figure}

For sufficiently high $\epsilon^c_\BM$, the equation $\Omega = j(\Omega)$ fails to yield a real, positive solution over a finite radial interval, producing discontinuities visible in the red curves of Fig.~\ref{fig:omega_of_r_profiles}. Although the code provides a numerical solution, such discontinuities imply that the model is not physically reliable and cannot be used as initial data for numerical relativity codes.

The impact on equilibrium sequences is minimal, as the discontinuities are confined to a few radial grid points near the equator. When computing integral quantities, such as the gravitational mass, the resulting error is negligible. Other field quantities involved in the integrals, such as pressure or energy density, exhibit jumps in the radial profile of up to 5\% relative to neighboring points in the worst observed case, affecting at most six radial grid points out of the 400 that fall inside the star. The number of affected points decreases rapidly in the angular direction, so that the discontinuity is completely smoothed out within a few angular grid points.

To verify that the discontinuities described above are not numerical artifacts, we performed a series of independent tests. First, we compared our one-fluid configurations directly against the equilibrium models reported by~\cite{10.1093/mnras/stab3565} in Table~6, which we treat as reference solutions. Figure~\ref{fig:Iosif_errors} summarizes the relative differences between the two sets of results, showing that our models reproduce their data with high accuracy: the gravitational mass differs by at most $\sim 1.3\%$ for configurations with $(\lambda_1,\,\lambda_2)=(1.6,\,1)$, while the equatorial radius and the quantities related to the angular velocity, including the central, equatorial, and maximum values, agree to within $0.4\%$.  This error on the mass can be traced back to differences in the implementation of the EOS, rather than to the numerical method itself. Direct comparison of the full angular velocity profiles on the equatorial plane with those shown in Fig.~12 of~\cite{10.1093/mnras/stab3565} confirms the agreement, with discrepancies well below $1\%$. Importantly, in all of these benchmark cases, our solutions for $\Omega(r)$ are continuous across the stellar interior, demonstrating that the observed discontinuities at high central energy densities and high angular momentum are not intrinsic to the numerical implementation.

We also increased the grid resolution in both the radial and angular directions, from $(800,\,400)$ to $(1600,\,800)$, and tightened the convergence tolerance from $10^{-9}$ to $10^{-12}$ (global quantities such as the gravitational mass already converge at the level of $10^{-7}$). Root finding within the algorithm was maintained at machine accuracy. At high resolution, small oscillatory features arise near the stellar surface due to the steep decrease of matter variables. They can however be removed by switching the interpolation scheme from cubic to linear.

The location and extent of the discontinuous radial interval remain unchanged within numerical error.

Together, these tests indicate that the discontinuities are not due to insufficient resolution or implementation errors, but rather reflect the absence of a real solution for $\Omega$ over the affected interval.


\begin{thebibliography}{76}%
\makeatletter
\providecommand \@ifxundefined [1]{%
 \@ifx{#1\undefined}
}%
\providecommand \@ifnum [1]{%
 \ifnum #1\expandafter \@firstoftwo
 \else \expandafter \@secondoftwo
 \fi
}%
\providecommand \@ifx [1]{%
 \ifx #1\expandafter \@firstoftwo
 \else \expandafter \@secondoftwo
 \fi
}%
\providecommand \natexlab [1]{#1}%
\providecommand \enquote  [1]{``#1''}%
\providecommand \bibnamefont  [1]{#1}%
\providecommand \bibfnamefont [1]{#1}%
\providecommand \citenamefont [1]{#1}%
\providecommand \href@noop [0]{\@secondoftwo}%
\providecommand \href [0]{\begingroup \@sanitize@url \@href}%
\providecommand \@href[1]{\@@startlink{#1}\@@href}%
\providecommand \@@href[1]{\endgroup#1\@@endlink}%
\providecommand \@sanitize@url [0]{\catcode `\\12\catcode `\$12\catcode `\&12\catcode `\#12\catcode `\^12\catcode `\_12\catcode `\%12\relax}%
\providecommand \@@startlink[1]{}%
\providecommand \@@endlink[0]{}%
\providecommand \url  [0]{\begingroup\@sanitize@url \@url }%
\providecommand \@url [1]{\endgroup\@href {#1}{\urlprefix }}%
\providecommand \urlprefix  [0]{URL }%
\providecommand \Eprint [0]{\href }%
\providecommand \doibase [0]{http://dx.doi.org/}%
\providecommand \selectlanguage [0]{\@gobble}%
\providecommand \bibinfo  [0]{\@secondoftwo}%
\providecommand \bibfield  [0]{\@secondoftwo}%
\providecommand \translation [1]{[#1]}%
\providecommand \BibitemOpen [0]{}%
\providecommand \bibitemStop [0]{}%
\providecommand \bibitemNoStop [0]{.\EOS\space}%
\providecommand \EOS [0]{\spacefactor3000\relax}%
\providecommand \BibitemShut  [1]{\csname bibitem#1\endcsname}%
\let\auto@bib@innerbib\@empty
\bibitem [{\citenamefont {Abbott}\ \emph {et~al.}(2017)\citenamefont {Abbott} \emph {et~al.}}]{LIGOScientific:2017vwq}%
  \BibitemOpen
  \bibfield  {author} {\bibinfo {author} {\bibfnamefont {B.~P.}\ \bibnamefont {Abbott}} \emph {et~al.} (\bibinfo {collaboration} {LIGO Scientific, Virgo}),\ }\href {\doibase 10.1103/PhysRevLett.119.161101} {\bibfield  {journal} {\bibinfo  {journal} {Phys. Rev. Lett.}\ }\textbf {\bibinfo {volume} {119}},\ \bibinfo {pages} {161101} (\bibinfo {year} {2017})},\ \Eprint {http://arxiv.org/abs/1710.05832} {arXiv:1710.05832 [gr-qc]} \BibitemShut {NoStop}%
\bibitem [{\citenamefont {Bauswein}\ \emph {et~al.}(2017)\citenamefont {Bauswein}, \citenamefont {Just}, \citenamefont {Janka},\ and\ \citenamefont {Stergioulas}}]{Bauswein:2017vtn}%
  \BibitemOpen
  \bibfield  {author} {\bibinfo {author} {\bibfnamefont {A.}~\bibnamefont {Bauswein}}, \bibinfo {author} {\bibfnamefont {O.}~\bibnamefont {Just}}, \bibinfo {author} {\bibfnamefont {H.-T.}\ \bibnamefont {Janka}}, \ and\ \bibinfo {author} {\bibfnamefont {N.}~\bibnamefont {Stergioulas}},\ }\href {\doibase 10.3847/2041-8213/aa9994} {\bibfield  {journal} {\bibinfo  {journal} {Astrophys. J. Lett.}\ }\textbf {\bibinfo {volume} {850}},\ \bibinfo {pages} {L34} (\bibinfo {year} {2017})},\ \Eprint {http://arxiv.org/abs/1710.06843} {arXiv:1710.06843 [astro-ph.HE]} \BibitemShut {NoStop}%
\bibitem [{\citenamefont {Radice}\ \emph {et~al.}(2018)\citenamefont {Radice}, \citenamefont {Perego}, \citenamefont {Zappa},\ and\ \citenamefont {Bernuzzi}}]{Radice:2017lry}%
  \BibitemOpen
  \bibfield  {author} {\bibinfo {author} {\bibfnamefont {D.}~\bibnamefont {Radice}}, \bibinfo {author} {\bibfnamefont {A.}~\bibnamefont {Perego}}, \bibinfo {author} {\bibfnamefont {F.}~\bibnamefont {Zappa}}, \ and\ \bibinfo {author} {\bibfnamefont {S.}~\bibnamefont {Bernuzzi}},\ }\href {\doibase 10.3847/2041-8213/aaa402} {\bibfield  {journal} {\bibinfo  {journal} {Astrophys. J. Lett.}\ }\textbf {\bibinfo {volume} {852}},\ \bibinfo {pages} {L29} (\bibinfo {year} {2018})},\ \Eprint {http://arxiv.org/abs/1711.03647} {arXiv:1711.03647 [astro-ph.HE]} \BibitemShut {NoStop}%
\bibitem [{\citenamefont {Shibata}\ and\ \citenamefont {Uryu}(2000)}]{Shibata:1999wm}%
  \BibitemOpen
  \bibfield  {author} {\bibinfo {author} {\bibfnamefont {M.}~\bibnamefont {Shibata}}\ and\ \bibinfo {author} {\bibfnamefont {K.}~\bibnamefont {Uryu}},\ }\href {\doibase 10.1103/PhysRevD.61.064001} {\bibfield  {journal} {\bibinfo  {journal} {Phys. Rev. D}\ }\textbf {\bibinfo {volume} {61}},\ \bibinfo {pages} {064001} (\bibinfo {year} {2000})},\ \Eprint {http://arxiv.org/abs/gr-qc/9911058} {arXiv:gr-qc/9911058} \BibitemShut {NoStop}%
\bibitem [{\citenamefont {Baiotti}\ and\ \citenamefont {Rezzolla}(2017)}]{Baiotti:2016qnr}%
  \BibitemOpen
  \bibfield  {author} {\bibinfo {author} {\bibfnamefont {L.}~\bibnamefont {Baiotti}}\ and\ \bibinfo {author} {\bibfnamefont {L.}~\bibnamefont {Rezzolla}},\ }\href {\doibase 10.1088/1361-6633/aa67bb} {\bibfield  {journal} {\bibinfo  {journal} {Rept. Prog. Phys.}\ }\textbf {\bibinfo {volume} {80}},\ \bibinfo {pages} {096901} (\bibinfo {year} {2017})},\ \Eprint {http://arxiv.org/abs/1607.03540} {arXiv:1607.03540 [gr-qc]} \BibitemShut {NoStop}%
\bibitem [{\citenamefont {Paschalidis}\ and\ \citenamefont {Stergioulas}(2017)}]{Paschalidis:2016vmz}%
  \BibitemOpen
  \bibfield  {author} {\bibinfo {author} {\bibfnamefont {V.}~\bibnamefont {Paschalidis}}\ and\ \bibinfo {author} {\bibfnamefont {N.}~\bibnamefont {Stergioulas}},\ }\href {\doibase 10.1007/s41114-017-0008-x} {\bibfield  {journal} {\bibinfo  {journal} {Living Rev. Rel.}\ }\textbf {\bibinfo {volume} {20}},\ \bibinfo {pages} {7} (\bibinfo {year} {2017})},\ \Eprint {http://arxiv.org/abs/1612.03050} {arXiv:1612.03050 [astro-ph.HE]} \BibitemShut {NoStop}%
\bibitem [{\citenamefont {Baumgarte}\ \emph {et~al.}(2000)\citenamefont {Baumgarte}, \citenamefont {Shapiro},\ and\ \citenamefont {Shibata}}]{Baumgarte:1999cq}%
  \BibitemOpen
  \bibfield  {author} {\bibinfo {author} {\bibfnamefont {T.~W.}\ \bibnamefont {Baumgarte}}, \bibinfo {author} {\bibfnamefont {S.~L.}\ \bibnamefont {Shapiro}}, \ and\ \bibinfo {author} {\bibfnamefont {M.}~\bibnamefont {Shibata}},\ }\href {\doibase 10.1086/312425} {\bibfield  {journal} {\bibinfo  {journal} {Astrophys. J. Lett.}\ }\textbf {\bibinfo {volume} {528}},\ \bibinfo {pages} {L29} (\bibinfo {year} {2000})},\ \Eprint {http://arxiv.org/abs/astro-ph/9910565} {arXiv:astro-ph/9910565} \BibitemShut {NoStop}%
\bibitem [{\citenamefont {Komatsu}\ \emph {et~al.}(1989{\natexlab{a}})\citenamefont {Komatsu}, \citenamefont {Eriguchi},\ and\ \citenamefont {Hachisu}}]{10.1093/mnras/239.1.153}%
  \BibitemOpen
  \bibfield  {author} {\bibinfo {author} {\bibfnamefont {H.}~\bibnamefont {Komatsu}}, \bibinfo {author} {\bibfnamefont {Y.}~\bibnamefont {Eriguchi}}, \ and\ \bibinfo {author} {\bibfnamefont {I.}~\bibnamefont {Hachisu}},\ }\href {\doibase 10.1093/mnras/239.1.153} {\bibfield  {journal} {\bibinfo  {journal} {MNRAS}\ }\textbf {\bibinfo {volume} {239}},\ \bibinfo {pages} {153} (\bibinfo {year} {1989}{\natexlab{a}})}\BibitemShut {NoStop}%
\bibitem [{\citenamefont {Kastaun}\ and\ \citenamefont {Galeazzi}(2015)}]{PhysRevD.91.064027}%
  \BibitemOpen
  \bibfield  {author} {\bibinfo {author} {\bibfnamefont {W.}~\bibnamefont {Kastaun}}\ and\ \bibinfo {author} {\bibfnamefont {F.}~\bibnamefont {Galeazzi}},\ }\href {\doibase 10.1103/PhysRevD.91.064027} {\bibfield  {journal} {\bibinfo  {journal} {Phys. Rev. D}\ }\textbf {\bibinfo {volume} {91}},\ \bibinfo {pages} {064027} (\bibinfo {year} {2015})}\BibitemShut {NoStop}%
\bibitem [{\citenamefont {Hanauske}\ \emph {et~al.}(2017)\citenamefont {Hanauske}, \citenamefont {Takami}, \citenamefont {Bovard}, \citenamefont {Rezzolla}, \citenamefont {Font}, \citenamefont {Galeazzi},\ and\ \citenamefont {St\"ocker}}]{PhysRevD.96.043004}%
  \BibitemOpen
  \bibfield  {author} {\bibinfo {author} {\bibfnamefont {M.}~\bibnamefont {Hanauske}}, \bibinfo {author} {\bibfnamefont {K.}~\bibnamefont {Takami}}, \bibinfo {author} {\bibfnamefont {L.}~\bibnamefont {Bovard}}, \bibinfo {author} {\bibfnamefont {L.}~\bibnamefont {Rezzolla}}, \bibinfo {author} {\bibfnamefont {J.~A.}\ \bibnamefont {Font}}, \bibinfo {author} {\bibfnamefont {F.}~\bibnamefont {Galeazzi}}, \ and\ \bibinfo {author} {\bibfnamefont {H.}~\bibnamefont {St\"ocker}},\ }\href {\doibase 10.1103/PhysRevD.96.043004} {\bibfield  {journal} {\bibinfo  {journal} {Phys. Rev. D}\ }\textbf {\bibinfo {volume} {96}},\ \bibinfo {pages} {043004} (\bibinfo {year} {2017})}\BibitemShut {NoStop}%
\bibitem [{\citenamefont {Ciolfi}\ \emph {et~al.}(2017)\citenamefont {Ciolfi}, \citenamefont {Kastaun}, \citenamefont {Giacomazzo}, \citenamefont {Endrizzi}, \citenamefont {Siegel},\ and\ \citenamefont {Perna}}]{PhysRevD.95.063016}%
  \BibitemOpen
  \bibfield  {author} {\bibinfo {author} {\bibfnamefont {R.}~\bibnamefont {Ciolfi}}, \bibinfo {author} {\bibfnamefont {W.}~\bibnamefont {Kastaun}}, \bibinfo {author} {\bibfnamefont {B.}~\bibnamefont {Giacomazzo}}, \bibinfo {author} {\bibfnamefont {A.}~\bibnamefont {Endrizzi}}, \bibinfo {author} {\bibfnamefont {D.~M.}\ \bibnamefont {Siegel}}, \ and\ \bibinfo {author} {\bibfnamefont {R.}~\bibnamefont {Perna}},\ }\href {\doibase 10.1103/PhysRevD.95.063016} {\bibfield  {journal} {\bibinfo  {journal} {Phys. Rev. D}\ }\textbf {\bibinfo {volume} {95}},\ \bibinfo {pages} {063016} (\bibinfo {year} {2017})}\BibitemShut {NoStop}%
\bibitem [{\citenamefont {De~Pietri}\ \emph {et~al.}(2018)\citenamefont {De~Pietri}, \citenamefont {Feo}, \citenamefont {Font}, \citenamefont {L\"offler}, \citenamefont {Maione}, \citenamefont {Pasquali},\ and\ \citenamefont {Stergioulas}}]{PhysRevLett.120.221101}%
  \BibitemOpen
  \bibfield  {author} {\bibinfo {author} {\bibfnamefont {R.}~\bibnamefont {De~Pietri}}, \bibinfo {author} {\bibfnamefont {A.}~\bibnamefont {Feo}}, \bibinfo {author} {\bibfnamefont {J.~A.}\ \bibnamefont {Font}}, \bibinfo {author} {\bibfnamefont {F.}~\bibnamefont {L\"offler}}, \bibinfo {author} {\bibfnamefont {F.}~\bibnamefont {Maione}}, \bibinfo {author} {\bibfnamefont {M.}~\bibnamefont {Pasquali}}, \ and\ \bibinfo {author} {\bibfnamefont {N.}~\bibnamefont {Stergioulas}},\ }\href {\doibase 10.1103/PhysRevLett.120.221101} {\bibfield  {journal} {\bibinfo  {journal} {Phys. Rev. Lett.}\ }\textbf {\bibinfo {volume} {120}},\ \bibinfo {pages} {221101} (\bibinfo {year} {2018})}\BibitemShut {NoStop}%
\bibitem [{\citenamefont {East}\ \emph {et~al.}(2019)\citenamefont {East}, \citenamefont {Paschalidis}, \citenamefont {Pretorius},\ and\ \citenamefont {Tsokaros}}]{PhysRevD.100.124042}%
  \BibitemOpen
  \bibfield  {author} {\bibinfo {author} {\bibfnamefont {W.~E.}\ \bibnamefont {East}}, \bibinfo {author} {\bibfnamefont {V.}~\bibnamefont {Paschalidis}}, \bibinfo {author} {\bibfnamefont {F.}~\bibnamefont {Pretorius}}, \ and\ \bibinfo {author} {\bibfnamefont {A.}~\bibnamefont {Tsokaros}},\ }\href {\doibase 10.1103/PhysRevD.100.124042} {\bibfield  {journal} {\bibinfo  {journal} {Phys. Rev. D}\ }\textbf {\bibinfo {volume} {100}},\ \bibinfo {pages} {124042} (\bibinfo {year} {2019})}\BibitemShut {NoStop}%
\bibitem [{\citenamefont {Takami}\ \emph {et~al.}(2015)\citenamefont {Takami}, \citenamefont {Rezzolla},\ and\ \citenamefont {Baiotti}}]{PhysRevD.91.064001}%
  \BibitemOpen
  \bibfield  {author} {\bibinfo {author} {\bibfnamefont {K.}~\bibnamefont {Takami}}, \bibinfo {author} {\bibfnamefont {L.}~\bibnamefont {Rezzolla}}, \ and\ \bibinfo {author} {\bibfnamefont {L.}~\bibnamefont {Baiotti}},\ }\href {\doibase 10.1103/PhysRevD.91.064001} {\bibfield  {journal} {\bibinfo  {journal} {Phys. Rev. D}\ }\textbf {\bibinfo {volume} {91}},\ \bibinfo {pages} {064001} (\bibinfo {year} {2015})}\BibitemShut {NoStop}%
\bibitem [{\citenamefont {Ury\={u}}\ \emph {et~al.}(2017)\citenamefont {Ury\={u}}, \citenamefont {Tsokaros}, \citenamefont {Baiotti}, \citenamefont {Galeazzi}, \citenamefont {Taniguchi},\ and\ \citenamefont {Yoshida}}]{Uryu:2017}%
  \BibitemOpen
  \bibfield  {author} {\bibinfo {author} {\bibfnamefont {K.}~\bibnamefont {Ury\={u}}}, \bibinfo {author} {\bibfnamefont {A.}~\bibnamefont {Tsokaros}}, \bibinfo {author} {\bibfnamefont {L.}~\bibnamefont {Baiotti}}, \bibinfo {author} {\bibfnamefont {F.}~\bibnamefont {Galeazzi}}, \bibinfo {author} {\bibfnamefont {K.}~\bibnamefont {Taniguchi}}, \ and\ \bibinfo {author} {\bibfnamefont {S.}~\bibnamefont {Yoshida}},\ }\href {\doibase 10.1103/PhysRevD.96.103011} {\bibfield  {journal} {\bibinfo  {journal} {Phys. Rev. D}\ }\textbf {\bibinfo {volume} {96}},\ \bibinfo {pages} {103011} (\bibinfo {year} {2017})}\BibitemShut {NoStop}%
\bibitem [{\citenamefont {Passamonti}\ and\ \citenamefont {Andersson}(2020)}]{10.1093/mnras/staa2725}%
  \BibitemOpen
  \bibfield  {author} {\bibinfo {author} {\bibfnamefont {A.}~\bibnamefont {Passamonti}}\ and\ \bibinfo {author} {\bibfnamefont {N.}~\bibnamefont {Andersson}},\ }\href {\doibase 10.1093/mnras/staa2725} {\bibfield  {journal} {\bibinfo  {journal} {Monthly Notices of the Royal Astronomical Society}\ }\textbf {\bibinfo {volume} {498}},\ \bibinfo {pages} {5904} (\bibinfo {year} {2020})}\BibitemShut {NoStop}%
\bibitem [{\citenamefont {Weih}\ \emph {et~al.}(2020)\citenamefont {Weih}, \citenamefont {Hanauske},\ and\ \citenamefont {Rezzolla}}]{PhysRevLett.124.171103}%
  \BibitemOpen
  \bibfield  {author} {\bibinfo {author} {\bibfnamefont {L.~R.}\ \bibnamefont {Weih}}, \bibinfo {author} {\bibfnamefont {M.}~\bibnamefont {Hanauske}}, \ and\ \bibinfo {author} {\bibfnamefont {L.}~\bibnamefont {Rezzolla}},\ }\href {\doibase 10.1103/PhysRevLett.124.171103} {\bibfield  {journal} {\bibinfo  {journal} {Phys. Rev. Lett.}\ }\textbf {\bibinfo {volume} {124}},\ \bibinfo {pages} {171103} (\bibinfo {year} {2020})}\BibitemShut {NoStop}%
\bibitem [{\citenamefont {Cassing}\ and\ \citenamefont {Rezzolla}(2024)}]{Cassing:2024dxp}%
  \BibitemOpen
  \bibfield  {author} {\bibinfo {author} {\bibfnamefont {M.}~\bibnamefont {Cassing}}\ and\ \bibinfo {author} {\bibfnamefont {L.}~\bibnamefont {Rezzolla}},\ }\href {\doibase 10.1093/mnras/stae1527} {\bibfield  {journal} {\bibinfo  {journal} {Mon. Not. Roy. Astron. Soc.}\ }\textbf {\bibinfo {volume} {532}},\ \bibinfo {pages} {945} (\bibinfo {year} {2024})},\ \Eprint {http://arxiv.org/abs/2405.06609} {arXiv:2405.06609 [gr-qc]} \BibitemShut {NoStop}%
\bibitem [{\citenamefont {Camelio}\ \emph {et~al.}(2021)\citenamefont {Camelio}, \citenamefont {Dietrich}, \citenamefont {Rosswog},\ and\ \citenamefont {Haskell}}]{PhysRevD.103.063014}%
  \BibitemOpen
  \bibfield  {author} {\bibinfo {author} {\bibfnamefont {G.}~\bibnamefont {Camelio}}, \bibinfo {author} {\bibfnamefont {T.}~\bibnamefont {Dietrich}}, \bibinfo {author} {\bibfnamefont {S.}~\bibnamefont {Rosswog}}, \ and\ \bibinfo {author} {\bibfnamefont {B.}~\bibnamefont {Haskell}},\ }\href {\doibase 10.1103/PhysRevD.103.063014} {\bibfield  {journal} {\bibinfo  {journal} {Phys. Rev. D}\ }\textbf {\bibinfo {volume} {103}},\ \bibinfo {pages} {063014} (\bibinfo {year} {2021})}\BibitemShut {NoStop}%
\bibitem [{\citenamefont {Iosif}\ and\ \citenamefont {Stergioulas}(2021{\natexlab{a}})}]{10.1093/mnras/stab392}%
  \BibitemOpen
  \bibfield  {author} {\bibinfo {author} {\bibfnamefont {P.}~\bibnamefont {Iosif}}\ and\ \bibinfo {author} {\bibfnamefont {N.}~\bibnamefont {Stergioulas}},\ }\href {\doibase 10.1093/mnras/stab392} {\bibfield  {journal} {\bibinfo  {journal} {Monthly Notices of the Royal Astronomical Society}\ }\textbf {\bibinfo {volume} {503}},\ \bibinfo {pages} {850} (\bibinfo {year} {2021}{\natexlab{a}})}\BibitemShut {NoStop}%
\bibitem [{\citenamefont {Iosif}\ and\ \citenamefont {Stergioulas}(2021{\natexlab{b}})}]{10.1093/mnras/stab3565}%
  \BibitemOpen
  \bibfield  {author} {\bibinfo {author} {\bibfnamefont {P.}~\bibnamefont {Iosif}}\ and\ \bibinfo {author} {\bibfnamefont {N.}~\bibnamefont {Stergioulas}},\ }\href {\doibase 10.1093/mnras/stab3565} {\bibfield  {journal} {\bibinfo  {journal} {Monthly Notices of the Royal Astronomical Society}\ }\textbf {\bibinfo {volume} {510}},\ \bibinfo {pages} {2948} (\bibinfo {year} {2021}{\natexlab{b}})}\BibitemShut {NoStop}%
\bibitem [{\citenamefont {Bramante}\ \emph {et~al.}(2014)\citenamefont {Bramante}, \citenamefont {Fukushima}, \citenamefont {Kumar},\ and\ \citenamefont {Stopnitzky}}]{PhysRevD.89.015010}%
  \BibitemOpen
  \bibfield  {author} {\bibinfo {author} {\bibfnamefont {J.}~\bibnamefont {Bramante}}, \bibinfo {author} {\bibfnamefont {K.}~\bibnamefont {Fukushima}}, \bibinfo {author} {\bibfnamefont {J.}~\bibnamefont {Kumar}}, \ and\ \bibinfo {author} {\bibfnamefont {E.}~\bibnamefont {Stopnitzky}},\ }\href {\doibase 10.1103/PhysRevD.89.015010} {\bibfield  {journal} {\bibinfo  {journal} {Phys. Rev. D}\ }\textbf {\bibinfo {volume} {89}},\ \bibinfo {pages} {015010} (\bibinfo {year} {2014})}\BibitemShut {NoStop}%
\bibitem [{\citenamefont {Bell}\ \emph {et~al.}(2021)\citenamefont {Bell}, \citenamefont {Busoni}, \citenamefont {Motta}, \citenamefont {Robles}, \citenamefont {Thomas},\ and\ \citenamefont {Virgato}}]{Bell:2020obw}%
  \BibitemOpen
  \bibfield  {author} {\bibinfo {author} {\bibfnamefont {N.~F.}\ \bibnamefont {Bell}}, \bibinfo {author} {\bibfnamefont {G.}~\bibnamefont {Busoni}}, \bibinfo {author} {\bibfnamefont {T.~F.}\ \bibnamefont {Motta}}, \bibinfo {author} {\bibfnamefont {S.}~\bibnamefont {Robles}}, \bibinfo {author} {\bibfnamefont {A.~W.}\ \bibnamefont {Thomas}}, \ and\ \bibinfo {author} {\bibfnamefont {M.}~\bibnamefont {Virgato}},\ }\href {\doibase 10.1103/PhysRevLett.127.111803} {\bibfield  {journal} {\bibinfo  {journal} {Phys. Rev. Lett.}\ }\textbf {\bibinfo {volume} {127}},\ \bibinfo {pages} {111803} (\bibinfo {year} {2021})},\ \Eprint {http://arxiv.org/abs/2012.08918} {arXiv:2012.08918 [hep-ph]} \BibitemShut {NoStop}%
\bibitem [{\citenamefont {Bell}\ \emph {et~al.}(2020)\citenamefont {Bell}, \citenamefont {Busoni}, \citenamefont {Robles},\ and\ \citenamefont {Virgato}}]{Bell:2020jou}%
  \BibitemOpen
  \bibfield  {author} {\bibinfo {author} {\bibfnamefont {N.~F.}\ \bibnamefont {Bell}}, \bibinfo {author} {\bibfnamefont {G.}~\bibnamefont {Busoni}}, \bibinfo {author} {\bibfnamefont {S.}~\bibnamefont {Robles}}, \ and\ \bibinfo {author} {\bibfnamefont {M.}~\bibnamefont {Virgato}},\ }\href {\doibase 10.1088/1475-7516/2020/09/028} {\bibfield  {journal} {\bibinfo  {journal} {JCAP}\ }\textbf {\bibinfo {volume} {09}},\ \bibinfo {pages} {028} (\bibinfo {year} {2020})},\ \Eprint {http://arxiv.org/abs/2004.14888} {arXiv:2004.14888 [hep-ph]} \BibitemShut {NoStop}%
\bibitem [{\citenamefont {Stref}\ and\ \citenamefont {Lavalle}(2017)}]{Stref:2016uzb}%
  \BibitemOpen
  \bibfield  {author} {\bibinfo {author} {\bibfnamefont {M.}~\bibnamefont {Stref}}\ and\ \bibinfo {author} {\bibfnamefont {J.}~\bibnamefont {Lavalle}},\ }\href {\doibase 10.1103/PhysRevD.95.063003} {\bibfield  {journal} {\bibinfo  {journal} {Phys. Rev. D}\ }\textbf {\bibinfo {volume} {95}},\ \bibinfo {pages} {063003} (\bibinfo {year} {2017})},\ \Eprint {http://arxiv.org/abs/1610.02233} {arXiv:1610.02233 [astro-ph.CO]} \BibitemShut {NoStop}%
\bibitem [{\citenamefont {Lacroix}(2018)}]{Lacroix:2018zmg}%
  \BibitemOpen
  \bibfield  {author} {\bibinfo {author} {\bibfnamefont {T.}~\bibnamefont {Lacroix}},\ }\href {\doibase 10.1051/0004-6361/201832652} {\bibfield  {journal} {\bibinfo  {journal} {Astron. Astrophys.}\ }\textbf {\bibinfo {volume} {619}},\ \bibinfo {pages} {A46} (\bibinfo {year} {2018})},\ \Eprint {http://arxiv.org/abs/1801.01308} {arXiv:1801.01308 [astro-ph.GA]} \BibitemShut {NoStop}%
\bibitem [{\citenamefont {Nelson}\ \emph {et~al.}(2019)\citenamefont {Nelson}, \citenamefont {Reddy},\ and\ \citenamefont {Zhou}}]{Nelson_2019}%
  \BibitemOpen
  \bibfield  {author} {\bibinfo {author} {\bibfnamefont {A.~E.}\ \bibnamefont {Nelson}}, \bibinfo {author} {\bibfnamefont {S.}~\bibnamefont {Reddy}}, \ and\ \bibinfo {author} {\bibfnamefont {D.}~\bibnamefont {Zhou}},\ }\href {\doibase 10.1088/1475-7516/2019/07/012} {\bibfield  {journal} {\bibinfo  {journal} {Journal of Cosmology and Astroparticle Physics}\ }\textbf {\bibinfo {volume} {2019}},\ \bibinfo {pages} {012} (\bibinfo {year} {2019})}\BibitemShut {NoStop}%
\bibitem [{\citenamefont {Ellis}\ \emph {et~al.}(2018)\citenamefont {Ellis}, \citenamefont {H\"utsi}, \citenamefont {Kannike}, \citenamefont {Marzola}, \citenamefont {Raidal},\ and\ \citenamefont {Vaskonen}}]{PhysRevD.97.123007}%
  \BibitemOpen
  \bibfield  {author} {\bibinfo {author} {\bibfnamefont {J.}~\bibnamefont {Ellis}}, \bibinfo {author} {\bibfnamefont {G.}~\bibnamefont {H\"utsi}}, \bibinfo {author} {\bibfnamefont {K.}~\bibnamefont {Kannike}}, \bibinfo {author} {\bibfnamefont {L.}~\bibnamefont {Marzola}}, \bibinfo {author} {\bibfnamefont {M.}~\bibnamefont {Raidal}}, \ and\ \bibinfo {author} {\bibfnamefont {V.}~\bibnamefont {Vaskonen}},\ }\href {\doibase 10.1103/PhysRevD.97.123007} {\bibfield  {journal} {\bibinfo  {journal} {Phys. Rev. D}\ }\textbf {\bibinfo {volume} {97}},\ \bibinfo {pages} {123007} (\bibinfo {year} {2018})}\BibitemShut {NoStop}%
\bibitem [{\citenamefont {Del~Popolo}\ and\ \citenamefont {Le~Delliou}(2021)}]{DelPopolo:2021bom}%
  \BibitemOpen
  \bibfield  {author} {\bibinfo {author} {\bibfnamefont {A.}~\bibnamefont {Del~Popolo}}\ and\ \bibinfo {author} {\bibfnamefont {M.}~\bibnamefont {Le~Delliou}},\ }\href {\doibase 10.3390/galaxies9040123} {\bibfield  {journal} {\bibinfo  {journal} {Galaxies}\ }\textbf {\bibinfo {volume} {9}},\ \bibinfo {pages} {123} (\bibinfo {year} {2021})},\ \Eprint {http://arxiv.org/abs/2209.14151} {arXiv:2209.14151 [astro-ph.CO]} \BibitemShut {NoStop}%
\bibitem [{\citenamefont {Su\'arez}\ \emph {et~al.}(2014)\citenamefont {Su\'arez}, \citenamefont {Robles},\ and\ \citenamefont {Matos}}]{Suarez:2013iw}%
  \BibitemOpen
  \bibfield  {author} {\bibinfo {author} {\bibfnamefont {A.}~\bibnamefont {Su\'arez}}, \bibinfo {author} {\bibfnamefont {V.~H.}\ \bibnamefont {Robles}}, \ and\ \bibinfo {author} {\bibfnamefont {T.}~\bibnamefont {Matos}},\ }\href {\doibase 10.1007/978-3-319-02063-1_9} {\bibfield  {journal} {\bibinfo  {journal} {Astrophys. Space Sci. Proc.}\ }\textbf {\bibinfo {volume} {38}},\ \bibinfo {pages} {107} (\bibinfo {year} {2014})},\ \Eprint {http://arxiv.org/abs/1302.0903} {arXiv:1302.0903 [astro-ph.CO]} \BibitemShut {NoStop}%
\bibitem [{\citenamefont {Hippert}\ \emph {et~al.}(2023)\citenamefont {Hippert}, \citenamefont {Dillingham}, \citenamefont {Tan}, \citenamefont {Curtin}, \citenamefont {Noronha-Hostler},\ and\ \citenamefont {Yunes}}]{Hippert:2022snq}%
  \BibitemOpen
  \bibfield  {author} {\bibinfo {author} {\bibfnamefont {M.}~\bibnamefont {Hippert}}, \bibinfo {author} {\bibfnamefont {E.}~\bibnamefont {Dillingham}}, \bibinfo {author} {\bibfnamefont {H.}~\bibnamefont {Tan}}, \bibinfo {author} {\bibfnamefont {D.}~\bibnamefont {Curtin}}, \bibinfo {author} {\bibfnamefont {J.}~\bibnamefont {Noronha-Hostler}}, \ and\ \bibinfo {author} {\bibfnamefont {N.}~\bibnamefont {Yunes}},\ }\href {\doibase 10.1103/PhysRevD.107.115028} {\bibfield  {journal} {\bibinfo  {journal} {Phys. Rev. D}\ }\textbf {\bibinfo {volume} {107}},\ \bibinfo {pages} {115028} (\bibinfo {year} {2023})},\ \Eprint {http://arxiv.org/abs/2211.08590} {arXiv:2211.08590 [astro-ph.HE]} \BibitemShut {NoStop}%
\bibitem [{\citenamefont {Liu}\ \emph {et~al.}(2024)\citenamefont {Liu}, \citenamefont {Wei}, \citenamefont {Li}, \citenamefont {Burgio}, \citenamefont {Das},\ and\ \citenamefont {Schulze}}]{Liu:2024rix}%
  \BibitemOpen
  \bibfield  {author} {\bibinfo {author} {\bibfnamefont {H.-M.}\ \bibnamefont {Liu}}, \bibinfo {author} {\bibfnamefont {J.-B.}\ \bibnamefont {Wei}}, \bibinfo {author} {\bibfnamefont {Z.-H.}\ \bibnamefont {Li}}, \bibinfo {author} {\bibfnamefont {G.~F.}\ \bibnamefont {Burgio}}, \bibinfo {author} {\bibfnamefont {H.~C.}\ \bibnamefont {Das}}, \ and\ \bibinfo {author} {\bibfnamefont {H.~J.}\ \bibnamefont {Schulze}},\ }\href {\doibase 10.1103/PhysRevD.110.023024} {\bibfield  {journal} {\bibinfo  {journal} {Phys. Rev. D}\ }\textbf {\bibinfo {volume} {110}},\ \bibinfo {pages} {023024} (\bibinfo {year} {2024})},\ \Eprint {http://arxiv.org/abs/2403.17024} {arXiv:2403.17024 [nucl-th]} \BibitemShut {NoStop}%
\bibitem [{\citenamefont {Koehn}\ \emph {et~al.}(2024)\citenamefont {Koehn}, \citenamefont {Giangrandi}, \citenamefont {Kunert}, \citenamefont {Somasundaram}, \citenamefont {Sagun},\ and\ \citenamefont {Dietrich}}]{Koehn:2024gal}%
  \BibitemOpen
  \bibfield  {author} {\bibinfo {author} {\bibfnamefont {H.}~\bibnamefont {Koehn}}, \bibinfo {author} {\bibfnamefont {E.}~\bibnamefont {Giangrandi}}, \bibinfo {author} {\bibfnamefont {N.}~\bibnamefont {Kunert}}, \bibinfo {author} {\bibfnamefont {R.}~\bibnamefont {Somasundaram}}, \bibinfo {author} {\bibfnamefont {V.}~\bibnamefont {Sagun}}, \ and\ \bibinfo {author} {\bibfnamefont {T.}~\bibnamefont {Dietrich}},\ }\href {\doibase 10.1103/PhysRevD.110.103033} {\bibfield  {journal} {\bibinfo  {journal} {Phys. Rev. D}\ }\textbf {\bibinfo {volume} {110}},\ \bibinfo {pages} {103033} (\bibinfo {year} {2024})},\ \Eprint {http://arxiv.org/abs/2408.14711} {arXiv:2408.14711 [astro-ph.HE]} \BibitemShut {NoStop}%
\bibitem [{\citenamefont {Bezares}\ \emph {et~al.}(2019)\citenamefont {Bezares}, \citenamefont {Vigan\`o},\ and\ \citenamefont {Palenzuela}}]{Bezares:2019jcb}%
  \BibitemOpen
  \bibfield  {author} {\bibinfo {author} {\bibfnamefont {M.}~\bibnamefont {Bezares}}, \bibinfo {author} {\bibfnamefont {D.}~\bibnamefont {Vigan\`o}}, \ and\ \bibinfo {author} {\bibfnamefont {C.}~\bibnamefont {Palenzuela}},\ }\href {\doibase 10.1103/PhysRevD.100.044049} {\bibfield  {journal} {\bibinfo  {journal} {Phys. Rev. D}\ }\textbf {\bibinfo {volume} {100}},\ \bibinfo {pages} {044049} (\bibinfo {year} {2019})},\ \Eprint {http://arxiv.org/abs/1905.08551} {arXiv:1905.08551 [gr-qc]} \BibitemShut {NoStop}%
\bibitem [{\citenamefont {Di~Giovanni}\ \emph {et~al.}(2022)\citenamefont {Di~Giovanni}, \citenamefont {Sanchis-Gual}, \citenamefont {Guerra}, \citenamefont {Miravet-Ten\'es}, \citenamefont {Cerd\'a-Dur\'an},\ and\ \citenamefont {Font}}]{DiGiovanni:2022mkn}%
  \BibitemOpen
  \bibfield  {author} {\bibinfo {author} {\bibfnamefont {F.}~\bibnamefont {Di~Giovanni}}, \bibinfo {author} {\bibfnamefont {N.}~\bibnamefont {Sanchis-Gual}}, \bibinfo {author} {\bibfnamefont {D.}~\bibnamefont {Guerra}}, \bibinfo {author} {\bibfnamefont {M.}~\bibnamefont {Miravet-Ten\'es}}, \bibinfo {author} {\bibfnamefont {P.}~\bibnamefont {Cerd\'a-Dur\'an}}, \ and\ \bibinfo {author} {\bibfnamefont {J.~A.}\ \bibnamefont {Font}},\ }\href {\doibase 10.1103/PhysRevD.106.044008} {\bibfield  {journal} {\bibinfo  {journal} {Phys. Rev. D}\ }\textbf {\bibinfo {volume} {106}},\ \bibinfo {pages} {044008} (\bibinfo {year} {2022})},\ \Eprint {http://arxiv.org/abs/2206.00977} {arXiv:2206.00977 [gr-qc]} \BibitemShut {NoStop}%
\bibitem [{\citenamefont {R\"uter}\ \emph {et~al.}(2023)\citenamefont {R\"uter}, \citenamefont {Sagun}, \citenamefont {Tichy},\ and\ \citenamefont {Dietrich}}]{Ruter:2023uzc}%
  \BibitemOpen
  \bibfield  {author} {\bibinfo {author} {\bibfnamefont {H.~R.}\ \bibnamefont {R\"uter}}, \bibinfo {author} {\bibfnamefont {V.}~\bibnamefont {Sagun}}, \bibinfo {author} {\bibfnamefont {W.}~\bibnamefont {Tichy}}, \ and\ \bibinfo {author} {\bibfnamefont {T.}~\bibnamefont {Dietrich}},\ }\href {\doibase 10.1103/PhysRevD.108.124080} {\bibfield  {journal} {\bibinfo  {journal} {Phys. Rev. D}\ }\textbf {\bibinfo {volume} {108}},\ \bibinfo {pages} {124080} (\bibinfo {year} {2023})},\ \Eprint {http://arxiv.org/abs/2301.03568} {arXiv:2301.03568 [gr-qc]} \BibitemShut {NoStop}%
\bibitem [{\citenamefont {Bauswein}\ \emph {et~al.}(2023)\citenamefont {Bauswein}, \citenamefont {Guo}, \citenamefont {Lien}, \citenamefont {Lin},\ and\ \citenamefont {Wu}}]{Bauswein:2020kor}%
  \BibitemOpen
  \bibfield  {author} {\bibinfo {author} {\bibfnamefont {A.}~\bibnamefont {Bauswein}}, \bibinfo {author} {\bibfnamefont {G.}~\bibnamefont {Guo}}, \bibinfo {author} {\bibfnamefont {J.-H.}\ \bibnamefont {Lien}}, \bibinfo {author} {\bibfnamefont {Y.-H.}\ \bibnamefont {Lin}}, \ and\ \bibinfo {author} {\bibfnamefont {M.-R.}\ \bibnamefont {Wu}},\ }\href {\doibase 10.1103/PhysRevD.107.083002} {\bibfield  {journal} {\bibinfo  {journal} {Phys. Rev. D}\ }\textbf {\bibinfo {volume} {107}},\ \bibinfo {pages} {083002} (\bibinfo {year} {2023})},\ \Eprint {http://arxiv.org/abs/2012.11908} {arXiv:2012.11908 [astro-ph.HE]} \BibitemShut {NoStop}%
\bibitem [{\citenamefont {Emma}\ \emph {et~al.}(2022)\citenamefont {Emma}, \citenamefont {Schianchi}, \citenamefont {Pannarale}, \citenamefont {Sagun},\ and\ \citenamefont {Dietrich}}]{Emma:2022xjs}%
  \BibitemOpen
  \bibfield  {author} {\bibinfo {author} {\bibfnamefont {M.}~\bibnamefont {Emma}}, \bibinfo {author} {\bibfnamefont {F.}~\bibnamefont {Schianchi}}, \bibinfo {author} {\bibfnamefont {F.}~\bibnamefont {Pannarale}}, \bibinfo {author} {\bibfnamefont {V.}~\bibnamefont {Sagun}}, \ and\ \bibinfo {author} {\bibfnamefont {T.}~\bibnamefont {Dietrich}},\ }\href {\doibase 10.3390/particles5030024} {\bibfield  {journal} {\bibinfo  {journal} {Particles}\ }\textbf {\bibinfo {volume} {5}},\ \bibinfo {pages} {273} (\bibinfo {year} {2022})},\ \Eprint {http://arxiv.org/abs/2206.10887} {arXiv:2206.10887 [gr-qc]} \BibitemShut {NoStop}%
\bibitem [{\citenamefont {Routaray}\ \emph {et~al.}(2025)\citenamefont {Routaray}, \citenamefont {Chakrawarty}, \citenamefont {Patra},\ and\ \citenamefont {Kumar}}]{Routaray:2024lni}%
  \BibitemOpen
  \bibfield  {author} {\bibinfo {author} {\bibfnamefont {P.}~\bibnamefont {Routaray}}, \bibinfo {author} {\bibfnamefont {A.}~\bibnamefont {Chakrawarty}}, \bibinfo {author} {\bibfnamefont {N.~K.}\ \bibnamefont {Patra}}, \ and\ \bibinfo {author} {\bibfnamefont {B.}~\bibnamefont {Kumar}},\ }\href {\doibase 10.1016/j.dark.2025.102093} {\bibfield  {journal} {\bibinfo  {journal} {Phys. Dark Univ.}\ }\textbf {\bibinfo {volume} {50}},\ \bibinfo {pages} {102093} (\bibinfo {year} {2025})},\ \Eprint {http://arxiv.org/abs/2409.02131} {arXiv:2409.02131 [astro-ph.HE]} \BibitemShut {NoStop}%
\bibitem [{\citenamefont {Cronin}\ \emph {et~al.}(2023)\citenamefont {Cronin}, \citenamefont {Zhang},\ and\ \citenamefont {Kain}}]{Cronin:2023xzc}%
  \BibitemOpen
  \bibfield  {author} {\bibinfo {author} {\bibfnamefont {J.}~\bibnamefont {Cronin}}, \bibinfo {author} {\bibfnamefont {X.}~\bibnamefont {Zhang}}, \ and\ \bibinfo {author} {\bibfnamefont {B.}~\bibnamefont {Kain}},\ }\href {\doibase 10.1103/PhysRevD.108.103016} {\bibfield  {journal} {\bibinfo  {journal} {Phys. Rev. D}\ }\textbf {\bibinfo {volume} {108}},\ \bibinfo {pages} {103016} (\bibinfo {year} {2023})},\ \Eprint {http://arxiv.org/abs/2311.07714} {arXiv:2311.07714 [gr-qc]} \BibitemShut {NoStop}%
\bibitem [{\citenamefont {Konstantinou}(2024)}]{Konstantinou:2024ynd}%
  \BibitemOpen
  \bibfield  {author} {\bibinfo {author} {\bibfnamefont {A.}~\bibnamefont {Konstantinou}},\ }\href {\doibase 10.3847/1538-4357/ad4701} {\bibfield  {journal} {\bibinfo  {journal} {Astrophys. J.}\ }\textbf {\bibinfo {volume} {968}},\ \bibinfo {pages} {83} (\bibinfo {year} {2024})},\ \Eprint {http://arxiv.org/abs/2405.01487} {arXiv:2405.01487 [astro-ph.HE]} \BibitemShut {NoStop}%
\bibitem [{\citenamefont {Mourelle}\ \emph {et~al.}(2024)\citenamefont {Mourelle}, \citenamefont {Adam}, \citenamefont {Calder\'on~Bustillo},\ and\ \citenamefont {Sanchis-Gual}}]{Mourelle:2024qgo}%
  \BibitemOpen
  \bibfield  {author} {\bibinfo {author} {\bibfnamefont {J.~C.}\ \bibnamefont {Mourelle}}, \bibinfo {author} {\bibfnamefont {C.}~\bibnamefont {Adam}}, \bibinfo {author} {\bibfnamefont {J.}~\bibnamefont {Calder\'on~Bustillo}}, \ and\ \bibinfo {author} {\bibfnamefont {N.}~\bibnamefont {Sanchis-Gual}},\ }\href {\doibase 10.1103/PhysRevD.110.123019} {\bibfield  {journal} {\bibinfo  {journal} {Phys. Rev. D}\ }\textbf {\bibinfo {volume} {110}},\ \bibinfo {pages} {123019} (\bibinfo {year} {2024})},\ \Eprint {http://arxiv.org/abs/2403.13052} {arXiv:2403.13052 [gr-qc]} \BibitemShut {NoStop}%
\bibitem [{\citenamefont {Cipriani}\ \emph {et~al.}(2025)\citenamefont {Cipriani}, \citenamefont {Giangrandi}, \citenamefont {Sagun}, \citenamefont {Doneva},\ and\ \citenamefont {Yazadjiev}}]{Cipriani:2025tga}%
  \BibitemOpen
  \bibfield  {author} {\bibinfo {author} {\bibfnamefont {L.}~\bibnamefont {Cipriani}}, \bibinfo {author} {\bibfnamefont {E.}~\bibnamefont {Giangrandi}}, \bibinfo {author} {\bibfnamefont {V.}~\bibnamefont {Sagun}}, \bibinfo {author} {\bibfnamefont {D.~D.}\ \bibnamefont {Doneva}}, \ and\ \bibinfo {author} {\bibfnamefont {S.~S.}\ \bibnamefont {Yazadjiev}},\ }\href {\doibase 10.1103/qcl7-m5kf} {\bibfield  {journal} {\bibinfo  {journal} {Phys. Rev. D}\ }\textbf {\bibinfo {volume} {111}},\ \bibinfo {pages} {123005} (\bibinfo {year} {2025})},\ \Eprint {http://arxiv.org/abs/2502.17948} {arXiv:2502.17948 [astro-ph.HE]} \BibitemShut {NoStop}%
\bibitem [{\citenamefont {Giangrandi}\ \emph {et~al.}(2025)\citenamefont {Giangrandi}, \citenamefont {Rueter}, \citenamefont {Kunert}, \citenamefont {Emma}, \citenamefont {Abac}, \citenamefont {Adhikari}, \citenamefont {Dietrich}, \citenamefont {Sagun}, \citenamefont {Tichy},\ and\ \citenamefont {Providencia}}]{Giangrandi:2025rko}%
  \BibitemOpen
  \bibfield  {author} {\bibinfo {author} {\bibfnamefont {E.}~\bibnamefont {Giangrandi}}, \bibinfo {author} {\bibfnamefont {H.}~\bibnamefont {Rueter}}, \bibinfo {author} {\bibfnamefont {N.}~\bibnamefont {Kunert}}, \bibinfo {author} {\bibfnamefont {M.}~\bibnamefont {Emma}}, \bibinfo {author} {\bibfnamefont {A.}~\bibnamefont {Abac}}, \bibinfo {author} {\bibfnamefont {A.}~\bibnamefont {Adhikari}}, \bibinfo {author} {\bibfnamefont {T.}~\bibnamefont {Dietrich}}, \bibinfo {author} {\bibfnamefont {V.}~\bibnamefont {Sagun}}, \bibinfo {author} {\bibfnamefont {W.}~\bibnamefont {Tichy}}, \ and\ \bibinfo {author} {\bibfnamefont {C.}~\bibnamefont {Providencia}},\ }\href@noop {} {\  (\bibinfo {year} {2025})},\ \Eprint {http://arxiv.org/abs/2504.20825} {arXiv:2504.20825 [astro-ph.HE]} \BibitemShut {NoStop}%
\bibitem [{\citenamefont {Abac}\ \emph {et~al.}(2025)\citenamefont {Abac} \emph {et~al.}}]{ET:2025xjr}%
  \BibitemOpen
  \bibfield  {author} {\bibinfo {author} {\bibfnamefont {A.}~\bibnamefont {Abac}} \emph {et~al.} (\bibinfo {collaboration} {ET}),\ }\href@noop {} {\  (\bibinfo {year} {2025})},\ \Eprint {http://arxiv.org/abs/2503.12263} {arXiv:2503.12263 [gr-qc]} \BibitemShut {NoStop}%
\bibitem [{\citenamefont {Bozzola}\ \emph {et~al.}(2018)\citenamefont {Bozzola}, \citenamefont {Stergioulas},\ and\ \citenamefont {Bauswein}}]{Bozzola:2017qbu}%
  \BibitemOpen
  \bibfield  {author} {\bibinfo {author} {\bibfnamefont {G.}~\bibnamefont {Bozzola}}, \bibinfo {author} {\bibfnamefont {N.}~\bibnamefont {Stergioulas}}, \ and\ \bibinfo {author} {\bibfnamefont {A.}~\bibnamefont {Bauswein}},\ }\href {\doibase 10.1093/mnras/stx3002} {\bibfield  {journal} {\bibinfo  {journal} {Mon. Not. Roy. Astron. Soc.}\ }\textbf {\bibinfo {volume} {474}},\ \bibinfo {pages} {3557} (\bibinfo {year} {2018})},\ \Eprint {http://arxiv.org/abs/1709.02787} {arXiv:1709.02787 [gr-qc]} \BibitemShut {NoStop}%
\bibitem [{\citenamefont {Weih}\ \emph {et~al.}(2018)\citenamefont {Weih}, \citenamefont {Most},\ and\ \citenamefont {Rezzolla}}]{Weih:2017mcw}%
  \BibitemOpen
  \bibfield  {author} {\bibinfo {author} {\bibfnamefont {L.~R.}\ \bibnamefont {Weih}}, \bibinfo {author} {\bibfnamefont {E.~R.}\ \bibnamefont {Most}}, \ and\ \bibinfo {author} {\bibfnamefont {L.}~\bibnamefont {Rezzolla}},\ }\href {\doibase 10.1093/mnrasl/slx178} {\bibfield  {journal} {\bibinfo  {journal} {Mon. Not. Roy. Astron. Soc.}\ }\textbf {\bibinfo {volume} {473}},\ \bibinfo {pages} {L126} (\bibinfo {year} {2018})},\ \Eprint {http://arxiv.org/abs/1709.06058} {arXiv:1709.06058 [gr-qc]} \BibitemShut {NoStop}%
\bibitem [{\citenamefont {Ciolfi}\ \emph {et~al.}(2021)\citenamefont {Ciolfi}, \citenamefont {Stratta}, \citenamefont {Branchesi},\ and\ \citenamefont {Gendre}}]{Ciolfi2021}%
  \BibitemOpen
  \bibfield  {author} {\bibinfo {author} {\bibfnamefont {R.}~\bibnamefont {Ciolfi}}, \bibinfo {author} {\bibfnamefont {G.}~\bibnamefont {Stratta}}, \bibinfo {author} {\bibfnamefont {M.}~\bibnamefont {Branchesi}}, \ and\ \bibinfo {author} {\bibfnamefont {B.}~\bibnamefont {Gendre}},\ }\href {\doibase 10.1007/s10686-021-09795-9} {\bibfield  {journal} {\bibinfo  {journal} {Experimental Astronomy}\ }\textbf {\bibinfo {volume} {52}},\ \bibinfo {pages} {245} (\bibinfo {year} {2021})}\BibitemShut {NoStop}%
\bibitem [{\citenamefont {Rosati}\ \emph {et~al.}(2021)\citenamefont {Rosati}, \citenamefont {Basa}, \citenamefont {Blain}, \citenamefont {Bozzo},\ and\ \citenamefont {Branchesi}}]{Rosati2021}%
  \BibitemOpen
  \bibfield  {author} {\bibinfo {author} {\bibfnamefont {P.}~\bibnamefont {Rosati}}, \bibinfo {author} {\bibfnamefont {S.}~\bibnamefont {Basa}}, \bibinfo {author} {\bibfnamefont {A.~W.}\ \bibnamefont {Blain}}, \bibinfo {author} {\bibfnamefont {E.}~\bibnamefont {Bozzo}}, \ and\ \bibinfo {author} {\bibfnamefont {M.}~\bibnamefont {Branchesi}},\ }\href {\doibase 10.1007/s10686-021-09764-2} {\bibfield  {journal} {\bibinfo  {journal} {Experimental Astronomy}\ }\textbf {\bibinfo {volume} {52}},\ \bibinfo {pages} {407} (\bibinfo {year} {2021})}\BibitemShut {NoStop}%
\bibitem [{\citenamefont {Stergioulas}\ and\ \citenamefont {Friedman}(1995)}]{Stergioulas:1994ea}%
  \BibitemOpen
  \bibfield  {author} {\bibinfo {author} {\bibfnamefont {N.}~\bibnamefont {Stergioulas}}\ and\ \bibinfo {author} {\bibfnamefont {J.~L.}\ \bibnamefont {Friedman}},\ }\href {\doibase 10.1086/175605} {\bibfield  {journal} {\bibinfo  {journal} {Astrophys. J.}\ }\textbf {\bibinfo {volume} {444}},\ \bibinfo {pages} {306} (\bibinfo {year} {1995})},\ \Eprint {http://arxiv.org/abs/astro-ph/9411032} {arXiv:astro-ph/9411032} \BibitemShut {NoStop}%
\bibitem [{\citenamefont {Stergioulas}\ \emph {et~al.}(2004)\citenamefont {Stergioulas}, \citenamefont {Apostolatos},\ and\ \citenamefont {Font}}]{Stergioulas:2003ep}%
  \BibitemOpen
  \bibfield  {author} {\bibinfo {author} {\bibfnamefont {N.}~\bibnamefont {Stergioulas}}, \bibinfo {author} {\bibfnamefont {T.~A.}\ \bibnamefont {Apostolatos}}, \ and\ \bibinfo {author} {\bibfnamefont {J.~A.}\ \bibnamefont {Font}},\ }\href {\doibase 10.1111/j.1365-2966.2004.07973.x} {\bibfield  {journal} {\bibinfo  {journal} {Mon. Not. Roy. Astron. Soc.}\ }\textbf {\bibinfo {volume} {352}},\ \bibinfo {pages} {1089} (\bibinfo {year} {2004})},\ \Eprint {http://arxiv.org/abs/astro-ph/0312648} {arXiv:astro-ph/0312648} \BibitemShut {NoStop}%
\bibitem [{\citenamefont {Henriques}\ \emph {et~al.}(1990)\citenamefont {Henriques}, \citenamefont {Liddle},\ and\ \citenamefont {Moorhouse}}]{Henriques:1989ez}%
  \BibitemOpen
  \bibfield  {author} {\bibinfo {author} {\bibfnamefont {A.~B.}\ \bibnamefont {Henriques}}, \bibinfo {author} {\bibfnamefont {A.~R.}\ \bibnamefont {Liddle}}, \ and\ \bibinfo {author} {\bibfnamefont {R.~G.}\ \bibnamefont {Moorhouse}},\ }\href {\doibase 10.1016/0550-3213(90)90514-E} {\bibfield  {journal} {\bibinfo  {journal} {Nucl. Phys. B}\ }\textbf {\bibinfo {volume} {337}},\ \bibinfo {pages} {737} (\bibinfo {year} {1990})}\BibitemShut {NoStop}%
\bibitem [{\citenamefont {Mourelle}\ \emph {et~al.}(2025)\citenamefont {Mourelle}, \citenamefont {Sanchis-Gual},\ and\ \citenamefont {Font}}]{Mourelle:2025}%
  \BibitemOpen
  \bibfield  {author} {\bibinfo {author} {\bibfnamefont {J.~C.}\ \bibnamefont {Mourelle}}, \bibinfo {author} {\bibfnamefont {N.}~\bibnamefont {Sanchis-Gual}}, \ and\ \bibinfo {author} {\bibfnamefont {J.~A.}\ \bibnamefont {Font}},\ }\href@noop {} {\enquote {\bibinfo {title} {Differentially rotating fermion-boson stars},}\ } (\bibinfo {year} {2025}),\ \bibinfo {note} {manuscript in preparation}\BibitemShut {NoStop}%
\bibitem [{\citenamefont {Komatsu}\ \emph {et~al.}(1989{\natexlab{b}})\citenamefont {Komatsu}, \citenamefont {Eriguchi},\ and\ \citenamefont {Hachisu}}]{Komatsu:1989zz}%
  \BibitemOpen
  \bibfield  {author} {\bibinfo {author} {\bibfnamefont {H.}~\bibnamefont {Komatsu}}, \bibinfo {author} {\bibfnamefont {Y.}~\bibnamefont {Eriguchi}}, \ and\ \bibinfo {author} {\bibfnamefont {I.}~\bibnamefont {Hachisu}},\ }\href {\doibase https://doi.org/10.1093/mnras/237.2.355} {\bibfield  {journal} {\bibinfo  {journal} {Mon. Not. Roy. Astron. Soc.}\ }\textbf {\bibinfo {volume} {237}},\ \bibinfo {pages} {355} (\bibinfo {year} {1989}{\natexlab{b}})}\BibitemShut {NoStop}%
\bibitem [{\citenamefont {{Cook}}\ \emph {et~al.}(1992)\citenamefont {{Cook}}, \citenamefont {{Shapiro}},\ and\ \citenamefont {{Teukolsky}}}]{Cook:1992}%
  \BibitemOpen
  \bibfield  {author} {\bibinfo {author} {\bibfnamefont {G.~B.}\ \bibnamefont {{Cook}}}, \bibinfo {author} {\bibfnamefont {S.~L.}\ \bibnamefont {{Shapiro}}}, \ and\ \bibinfo {author} {\bibfnamefont {S.~A.}\ \bibnamefont {{Teukolsky}}},\ }\href {\doibase 10.1086/171849} {\bibfield  {journal} {\bibinfo  {journal} {\apj}\ }\textbf {\bibinfo {volume} {398}},\ \bibinfo {pages} {203} (\bibinfo {year} {1992})}\BibitemShut {NoStop}%
\bibitem [{\citenamefont {Villain}\ \emph {et~al.}(2004)\citenamefont {Villain}, \citenamefont {Pons}, \citenamefont {Cerda-Duran},\ and\ \citenamefont {Gourgoulhon}}]{Villain:2003ey}%
  \BibitemOpen
  \bibfield  {author} {\bibinfo {author} {\bibfnamefont {L.}~\bibnamefont {Villain}}, \bibinfo {author} {\bibfnamefont {J.~A.}\ \bibnamefont {Pons}}, \bibinfo {author} {\bibfnamefont {P.}~\bibnamefont {Cerda-Duran}}, \ and\ \bibinfo {author} {\bibfnamefont {E.}~\bibnamefont {Gourgoulhon}},\ }\href {\doibase 10.1051/0004-6361:20035619} {\bibfield  {journal} {\bibinfo  {journal} {Astron. Astrophys.}\ }\textbf {\bibinfo {volume} {418}},\ \bibinfo {pages} {283} (\bibinfo {year} {2004})},\ \Eprint {http://arxiv.org/abs/astro-ph/0310875} {arXiv:astro-ph/0310875} \BibitemShut {NoStop}%
\bibitem [{\citenamefont {Iosif}\ and\ \citenamefont {Stergioulas}(2021{\natexlab{c}})}]{Iosif:2021-09312}%
  \BibitemOpen
  \bibfield  {author} {\bibinfo {author} {\bibfnamefont {P.}~\bibnamefont {Iosif}}\ and\ \bibinfo {author} {\bibfnamefont {N.}~\bibnamefont {Stergioulas}},\ }\href {\doibase 10.3390/ECU2021-09312} {\bibfield  {journal} {\bibinfo  {journal} {Physical Sciences Forum}\ }\textbf {\bibinfo {volume} {2}} (\bibinfo {year} {2021}{\natexlab{c}}),\ 10.3390/ECU2021-09312}\BibitemShut {NoStop}%
\bibitem [{\citenamefont {Ansorg}\ \emph {et~al.}(2009)\citenamefont {Ansorg}, \citenamefont {Gondek-Rosińska},\ and\ \citenamefont {Villain}}]{Ansorg:2009}%
  \BibitemOpen
  \bibfield  {author} {\bibinfo {author} {\bibfnamefont {M.}~\bibnamefont {Ansorg}}, \bibinfo {author} {\bibfnamefont {D.}~\bibnamefont {Gondek-Rosińska}}, \ and\ \bibinfo {author} {\bibfnamefont {L.}~\bibnamefont {Villain}},\ }\href {\doibase 10.1111/j.1365-2966.2009.14904.x} {\bibfield  {journal} {\bibinfo  {journal} {Monthly Notices of the Royal Astronomical Society}\ }\textbf {\bibinfo {volume} {396}},\ \bibinfo {pages} {2359} (\bibinfo {year} {2009})}\BibitemShut {NoStop}%
\bibitem [{\citenamefont {{Gondek-Rosi{\'n}ska}}\ \emph {et~al.}(2017)\citenamefont {{Gondek-Rosi{\'n}ska}}, \citenamefont {{Kowalska}}, \citenamefont {{Villain}}, \citenamefont {{Ansorg}},\ and\ \citenamefont {{Kucaba}}}]{2017ApJ...837...58G}%
  \BibitemOpen
  \bibfield  {author} {\bibinfo {author} {\bibfnamefont {D.}~\bibnamefont {{Gondek-Rosi{\'n}ska}}}, \bibinfo {author} {\bibfnamefont {I.}~\bibnamefont {{Kowalska}}}, \bibinfo {author} {\bibfnamefont {L.}~\bibnamefont {{Villain}}}, \bibinfo {author} {\bibfnamefont {M.}~\bibnamefont {{Ansorg}}}, \ and\ \bibinfo {author} {\bibfnamefont {M.}~\bibnamefont {{Kucaba}}},\ }\href {\doibase 10.3847/1538-4357/aa56c1} {\bibfield  {journal} {\bibinfo  {journal} {\apj}\ }\textbf {\bibinfo {volume} {837}},\ \bibinfo {eid} {58} (\bibinfo {year} {2017})},\ \Eprint {http://arxiv.org/abs/1609.02336} {arXiv:1609.02336 [astro-ph.HE]} \BibitemShut {NoStop}%
\bibitem [{\citenamefont {Typel}\ \emph {et~al.}(2010)\citenamefont {Typel}, \citenamefont {R\"opke}, \citenamefont {Kl\"ahn}, \citenamefont {Blaschke},\ and\ \citenamefont {Wolter}}]{PhysRevC.81.015803}%
  \BibitemOpen
  \bibfield  {author} {\bibinfo {author} {\bibfnamefont {S.}~\bibnamefont {Typel}}, \bibinfo {author} {\bibfnamefont {G.}~\bibnamefont {R\"opke}}, \bibinfo {author} {\bibfnamefont {T.}~\bibnamefont {Kl\"ahn}}, \bibinfo {author} {\bibfnamefont {D.}~\bibnamefont {Blaschke}}, \ and\ \bibinfo {author} {\bibfnamefont {H.~H.}\ \bibnamefont {Wolter}},\ }\href {\doibase 10.1103/PhysRevC.81.015803} {\bibfield  {journal} {\bibinfo  {journal} {Phys. Rev. C}\ }\textbf {\bibinfo {volume} {81}},\ \bibinfo {pages} {015803} (\bibinfo {year} {2010})}\BibitemShut {NoStop}%
\bibitem [{\citenamefont {Antoniadis}\ \emph {et~al.}(2013)\citenamefont {Antoniadis} \emph {et~al.}}]{Antoniadis:2013pzd}%
  \BibitemOpen
  \bibfield  {author} {\bibinfo {author} {\bibfnamefont {J.}~\bibnamefont {Antoniadis}} \emph {et~al.},\ }\href {\doibase 10.1126/science.1233232} {\bibfield  {journal} {\bibinfo  {journal} {Science}\ }\textbf {\bibinfo {volume} {340}},\ \bibinfo {pages} {6131} (\bibinfo {year} {2013})},\ \Eprint {http://arxiv.org/abs/1304.6875} {arXiv:1304.6875 [astro-ph.HE]} \BibitemShut {NoStop}%
\bibitem [{\citenamefont {Romani}\ \emph {et~al.}(2021)\citenamefont {Romani}, \citenamefont {Kandel}, \citenamefont {Filippenko}, \citenamefont {Brink},\ and\ \citenamefont {Zheng}}]{Romani:2021xmb}%
  \BibitemOpen
  \bibfield  {author} {\bibinfo {author} {\bibfnamefont {R.~W.}\ \bibnamefont {Romani}}, \bibinfo {author} {\bibfnamefont {D.}~\bibnamefont {Kandel}}, \bibinfo {author} {\bibfnamefont {A.~V.}\ \bibnamefont {Filippenko}}, \bibinfo {author} {\bibfnamefont {T.~G.}\ \bibnamefont {Brink}}, \ and\ \bibinfo {author} {\bibfnamefont {W.}~\bibnamefont {Zheng}},\ }\href {\doibase 10.3847/2041-8213/abe2b4} {\bibfield  {journal} {\bibinfo  {journal} {Astrophys. J. Lett.}\ }\textbf {\bibinfo {volume} {908}},\ \bibinfo {pages} {L46} (\bibinfo {year} {2021})},\ \Eprint {http://arxiv.org/abs/2101.09822} {arXiv:2101.09822 [astro-ph.HE]} \BibitemShut {NoStop}%
\bibitem [{\citenamefont {Romani}\ \emph {et~al.}(2022)\citenamefont {Romani}, \citenamefont {Kandel}, \citenamefont {Filippenko}, \citenamefont {Brink},\ and\ \citenamefont {Zheng}}]{Romani:2022jhd}%
  \BibitemOpen
  \bibfield  {author} {\bibinfo {author} {\bibfnamefont {R.~W.}\ \bibnamefont {Romani}}, \bibinfo {author} {\bibfnamefont {D.}~\bibnamefont {Kandel}}, \bibinfo {author} {\bibfnamefont {A.~V.}\ \bibnamefont {Filippenko}}, \bibinfo {author} {\bibfnamefont {T.~G.}\ \bibnamefont {Brink}}, \ and\ \bibinfo {author} {\bibfnamefont {W.}~\bibnamefont {Zheng}},\ }\href {\doibase 10.3847/2041-8213/ac8007} {\bibfield  {journal} {\bibinfo  {journal} {Astrophys. J. Lett.}\ }\textbf {\bibinfo {volume} {934}},\ \bibinfo {pages} {L17} (\bibinfo {year} {2022})},\ \Eprint {http://arxiv.org/abs/2207.05124} {arXiv:2207.05124 [astro-ph.HE]} \BibitemShut {NoStop}%
\bibitem [{\citenamefont {Abbott}\ \emph {et~al.}(2018)\citenamefont {Abbott} \emph {et~al.}}]{LIGOScientific:2018cki}%
  \BibitemOpen
  \bibfield  {author} {\bibinfo {author} {\bibfnamefont {B.~P.}\ \bibnamefont {Abbott}} \emph {et~al.} (\bibinfo {collaboration} {LIGO Scientific, Virgo}),\ }\href {\doibase 10.1103/PhysRevLett.121.161101} {\bibfield  {journal} {\bibinfo  {journal} {Phys. Rev. Lett.}\ }\textbf {\bibinfo {volume} {121}},\ \bibinfo {pages} {161101} (\bibinfo {year} {2018})},\ \Eprint {http://arxiv.org/abs/1805.11581} {arXiv:1805.11581 [gr-qc]} \BibitemShut {NoStop}%
\bibitem [{\citenamefont {Abbott}\ \emph {et~al.}(2020)\citenamefont {Abbott} \emph {et~al.}}]{LIGOScientific:2020aai}%
  \BibitemOpen
  \bibfield  {author} {\bibinfo {author} {\bibfnamefont {B.~P.}\ \bibnamefont {Abbott}} \emph {et~al.} (\bibinfo {collaboration} {LIGO Scientific, Virgo}),\ }\href {\doibase 10.3847/2041-8213/ab75f5} {\bibfield  {journal} {\bibinfo  {journal} {Astrophys. J. Lett.}\ }\textbf {\bibinfo {volume} {892}},\ \bibinfo {pages} {L3} (\bibinfo {year} {2020})},\ \Eprint {http://arxiv.org/abs/2001.01761} {arXiv:2001.01761 [astro-ph.HE]} \BibitemShut {NoStop}%
\bibitem [{\citenamefont {Miller}\ \emph {et~al.}(2019)\citenamefont {Miller} \emph {et~al.}}]{Miller:2019cac}%
  \BibitemOpen
  \bibfield  {author} {\bibinfo {author} {\bibfnamefont {M.~C.}\ \bibnamefont {Miller}} \emph {et~al.},\ }\href {\doibase 10.3847/2041-8213/ab50c5} {\bibfield  {journal} {\bibinfo  {journal} {Astrophys. J. Lett.}\ }\textbf {\bibinfo {volume} {887}},\ \bibinfo {pages} {L24} (\bibinfo {year} {2019})},\ \Eprint {http://arxiv.org/abs/1912.05705} {arXiv:1912.05705 [astro-ph.HE]} \BibitemShut {NoStop}%
\bibitem [{\citenamefont {Riley}\ \emph {et~al.}(2019)\citenamefont {Riley} \emph {et~al.}}]{Riley:2019yda}%
  \BibitemOpen
  \bibfield  {author} {\bibinfo {author} {\bibfnamefont {T.~E.}\ \bibnamefont {Riley}} \emph {et~al.},\ }\href {\doibase 10.3847/2041-8213/ab481c} {\bibfield  {journal} {\bibinfo  {journal} {Astrophys. J. Lett.}\ }\textbf {\bibinfo {volume} {887}},\ \bibinfo {pages} {L21} (\bibinfo {year} {2019})},\ \Eprint {http://arxiv.org/abs/1912.05702} {arXiv:1912.05702 [astro-ph.HE]} \BibitemShut {NoStop}%
\bibitem [{\citenamefont {Miller}\ \emph {et~al.}(2021)\citenamefont {Miller} \emph {et~al.}}]{Miller:2021qha}%
  \BibitemOpen
  \bibfield  {author} {\bibinfo {author} {\bibfnamefont {M.~C.}\ \bibnamefont {Miller}} \emph {et~al.},\ }\href {\doibase 10.3847/2041-8213/ac089b} {\bibfield  {journal} {\bibinfo  {journal} {Astrophys. J. Lett.}\ }\textbf {\bibinfo {volume} {918}},\ \bibinfo {pages} {L28} (\bibinfo {year} {2021})},\ \Eprint {http://arxiv.org/abs/2105.06979} {arXiv:2105.06979 [astro-ph.HE]} \BibitemShut {NoStop}%
\bibitem [{\citenamefont {Riley}\ \emph {et~al.}(2021)\citenamefont {Riley} \emph {et~al.}}]{Riley:2021pdl}%
  \BibitemOpen
  \bibfield  {author} {\bibinfo {author} {\bibfnamefont {T.~E.}\ \bibnamefont {Riley}} \emph {et~al.},\ }\href {\doibase 10.3847/2041-8213/ac0a81} {\bibfield  {journal} {\bibinfo  {journal} {Astrophys. J. Lett.}\ }\textbf {\bibinfo {volume} {918}},\ \bibinfo {pages} {L27} (\bibinfo {year} {2021})},\ \Eprint {http://arxiv.org/abs/2105.06980} {arXiv:2105.06980 [astro-ph.HE]} \BibitemShut {NoStop}%
\bibitem [{\citenamefont {Choudhury}\ \emph {et~al.}(2024)\citenamefont {Choudhury} \emph {et~al.}}]{Choudhury:2024xbk}%
  \BibitemOpen
  \bibfield  {author} {\bibinfo {author} {\bibfnamefont {D.}~\bibnamefont {Choudhury}} \emph {et~al.},\ }\href {\doibase 10.3847/2041-8213/ad5a6f} {\bibfield  {journal} {\bibinfo  {journal} {Astrophys. J. Lett.}\ }\textbf {\bibinfo {volume} {971}},\ \bibinfo {pages} {L20} (\bibinfo {year} {2024})},\ \Eprint {http://arxiv.org/abs/2407.06789} {arXiv:2407.06789 [astro-ph.HE]} \BibitemShut {NoStop}%
\bibitem [{\citenamefont {Typel}(2018)}]{Typel:2018wmm}%
  \BibitemOpen
  \bibfield  {author} {\bibinfo {author} {\bibfnamefont {S.}~\bibnamefont {Typel}},\ }\href {\doibase 10.1088/1361-6471/aadea5} {\bibfield  {journal} {\bibinfo  {journal} {J. Phys. G}\ }\textbf {\bibinfo {volume} {45}},\ \bibinfo {pages} {114001} (\bibinfo {year} {2018})}\BibitemShut {NoStop}%
\bibitem [{\citenamefont {Colpi}\ \emph {et~al.}(1986)\citenamefont {Colpi}, \citenamefont {Shapiro},\ and\ \citenamefont {Wasserman}}]{Colpi:1986ye}%
  \BibitemOpen
  \bibfield  {author} {\bibinfo {author} {\bibfnamefont {M.}~\bibnamefont {Colpi}}, \bibinfo {author} {\bibfnamefont {S.~L.}\ \bibnamefont {Shapiro}}, \ and\ \bibinfo {author} {\bibfnamefont {I.}~\bibnamefont {Wasserman}},\ }\href {\doibase 10.1103/PhysRevLett.57.2485} {\bibfield  {journal} {\bibinfo  {journal} {Phys. Rev. Lett.}\ }\textbf {\bibinfo {volume} {57}},\ \bibinfo {pages} {2485} (\bibinfo {year} {1986})}\BibitemShut {NoStop}%
\bibitem [{\citenamefont {Karkevandi}\ \emph {et~al.}(2022)\citenamefont {Karkevandi}, \citenamefont {Shakeri}, \citenamefont {Sagun},\ and\ \citenamefont {Ivanytskyi}}]{Karkevandi:2021ygv}%
  \BibitemOpen
  \bibfield  {author} {\bibinfo {author} {\bibfnamefont {D.~R.}\ \bibnamefont {Karkevandi}}, \bibinfo {author} {\bibfnamefont {S.}~\bibnamefont {Shakeri}}, \bibinfo {author} {\bibfnamefont {V.}~\bibnamefont {Sagun}}, \ and\ \bibinfo {author} {\bibfnamefont {O.}~\bibnamefont {Ivanytskyi}},\ }\href {\doibase 10.1103/PhysRevD.105.023001} {\bibfield  {journal} {\bibinfo  {journal} {Phys. Rev. D}\ }\textbf {\bibinfo {volume} {105}},\ \bibinfo {pages} {023001} (\bibinfo {year} {2022})},\ \Eprint {http://arxiv.org/abs/2109.03801} {arXiv:2109.03801 [astro-ph.HE]} \BibitemShut {NoStop}%
\bibitem [{\citenamefont {Rafiei~Karkevandi}\ \emph {et~al.}(2021)\citenamefont {Rafiei~Karkevandi}, \citenamefont {Shakeri}, \citenamefont {Sagun},\ and\ \citenamefont {Ivanytskyi}}]{RafieiKarkevandi:2021hcc}%
  \BibitemOpen
  \bibfield  {author} {\bibinfo {author} {\bibfnamefont {D.}~\bibnamefont {Rafiei~Karkevandi}}, \bibinfo {author} {\bibfnamefont {S.}~\bibnamefont {Shakeri}}, \bibinfo {author} {\bibfnamefont {V.}~\bibnamefont {Sagun}}, \ and\ \bibinfo {author} {\bibfnamefont {O.}~\bibnamefont {Ivanytskyi}},\ }in\ \href {\doibase 10.1142/9789811269776_0307} {\emph {\bibinfo {booktitle} {{16th Marcel Grossmann Meeting on~Recent Developments in Theoretical and Experimental General Relativity, Astrophysics and Relativistic Field Theories}}}}\ (\bibinfo {year} {2021})\ \Eprint {http://arxiv.org/abs/2112.14231} {arXiv:2112.14231 [astro-ph.HE]} \BibitemShut {NoStop}%
\bibitem [{\citenamefont {Friedman}\ and\ \citenamefont {Stergioulas}(2013)}]{Friedman:2013xza}%
  \BibitemOpen
  \bibfield  {author} {\bibinfo {author} {\bibfnamefont {J.~L.}\ \bibnamefont {Friedman}}\ and\ \bibinfo {author} {\bibfnamefont {N.}~\bibnamefont {Stergioulas}},\ }\href {\doibase 10.1017/CBO9780511977596} {\emph {\bibinfo {title} {{Rotating Relativistic Stars}}}},\ Cambridge Monographs on Mathematical Physics\ (\bibinfo  {publisher} {Cambridge University Press},\ \bibinfo {year} {2013})\BibitemShut {NoStop}%
\bibitem [{\citenamefont {{De Pietri}}\ \emph {et~al.}(2020)\citenamefont {{De Pietri}}, \citenamefont {{Feo}}, \citenamefont {{Font}}, \citenamefont {{L{\"o}ffler}}, \citenamefont {{Pasquali}},\ and\ \citenamefont {{Stergioulas}}}]{2020PhRvD.101f4052D}%
  \BibitemOpen
  \bibfield  {author} {\bibinfo {author} {\bibfnamefont {R.}~\bibnamefont {{De Pietri}}}, \bibinfo {author} {\bibfnamefont {A.}~\bibnamefont {{Feo}}}, \bibinfo {author} {\bibfnamefont {J.~A.}\ \bibnamefont {{Font}}}, \bibinfo {author} {\bibfnamefont {F.}~\bibnamefont {{L{\"o}ffler}}}, \bibinfo {author} {\bibfnamefont {M.}~\bibnamefont {{Pasquali}}}, \ and\ \bibinfo {author} {\bibfnamefont {N.}~\bibnamefont {{Stergioulas}}},\ }\href {\doibase 10.1103/PhysRevD.101.064052} {\bibfield  {journal} {\bibinfo  {journal} {\prd}\ }\textbf {\bibinfo {volume} {101}},\ \bibinfo {eid} {064052} (\bibinfo {year} {2020})},\ \Eprint {http://arxiv.org/abs/1910.04036} {arXiv:1910.04036 [gr-qc]} \BibitemShut {NoStop}%
\end{thebibliography}

%

\end{document}